\documentclass[twocolumn,floats,floatfix,showpacs,prd,superscriptaddress,nofootinbib]{revtex4-1}

\usepackage{graphicx,epsfig}
\usepackage{amssymb,amsmath,amsthm,amsfonts}
\usepackage{bm} 

\usepackage[inline]{enumitem}
\usepackage[linktocpage]{hyperref}
\usepackage[caption=false]{subfig}
\usepackage[usenames]{color}
\usepackage{url}
\usepackage[inline]{enumitem}

\def\p{\partial}
\def\dif{{\rm{d}}}
\def\McZ{M_{\rm c,0}}
\def\MZ{M_{\rm 0}}
\def\chiZ{\chi_{\rm 0}}
\def\tI{t_{\rm I}}

\def\MII{M_{\rm II}}
\def\chiII{\chi_{\rm II}}
\def\McII{M_{\rm c,II}}
\def\Z{\mathcal{Z}}

\begin{document}
\title{\large Evolution of black hole shadows from superradiance}

\author{Gast\'{o}n Creci}\email{g.f.crecikeinbaum@uu.com}
\author{Stefan Vandoren}\email{s.j.g.vandoren@uu.nl}
\affiliation{Institute for Theoretical Physics, Utrecht University, Princetonplein 5, 3584 CE Utrecht, The Netherlands}
\author{Helvi Witek}\email{hwitek@illinois.edu}
\affiliation{Department of Physics, University of Illinois at Urbana-Champaign, Urbana, Illinois 61801, USA}
\affiliation{Department of Physics, King's College London, Strand, London, WC2R 2LS, United Kingdom}

\begin{abstract} 
Black holes have turned into cosmic laboratories to search for ultralight scalars by virtue of the superradiant instability.
In this paper we present a detailed study of the impact of the superradiant evolution on the black hole shadow
and investigate the exciting possibility to explore it with future observations of very long baseline interferometry.
We simulated the superradiant evolution numerically, in the adiabatic regime, and derived analytic approximations modeling the process.
Driven by superradiance, we evolve the black hole shadow diameter and
\begin{enumerate*}[label={(\roman*)}]
\item find that it can change by a few $\mu$as, just below the current resolution of the Event Horizon Telescope,
        albeit on timescales that are longer than realistic observation times;
\item show that the shadow diameter can either shrink or grow;
and
\item explore in detail how the shadow's end state is determined by the initial parameters and coupling.
\end{enumerate*}
\end{abstract}


\maketitle

\section{Introduction and main results}
We have entered an exciting era in gravitational physics 
in which we observe black holes 
-- among the most fascinating predictions of Einstein's theory of gravity --
on a regular basis.
These observations range from the ``drumbeat'' of stellar-mass black hole collisions as detected by ground-based gravitational-wave detectors~\cite{Abbott:2016blz,
TheLIGOScientific:2016pea,LIGOScientific:2018mvr}~{\footnote{The gravitational-wave detector network has now reached truly 
global scales with the LIGO detectors in the United States, Virgo in Italy, and, 
as of the beginning of March 2020, KAGRA in Japan.}}
to the impressive images of the shadow of the supermassive black hole lurking at the center of galaxy M87
taken by the Event Horizon Telescope (EHT)~\cite{Akiyama:2019cqa,Akiyama:2019eap}.

They have opened up novel paths to address long-standing puzzles in modern physics. 
The present paper revolves around the exciting possibility to employ black holes as cosmic particle detectors
for ultralight fundamental fields~\cite{Arvanitaki:2009fg,Arvanitaki:2010sy,Brito:2014wla,Baumann:2018vus}
some of which have become popular dark matter candidates~\cite{Hui:2016ltb}.
This is possible because of the superradiant instability of black holes~\footnote{The effect of superradiance is, in fact, not owned by black holes: it has first been conceived by Zel’dovich~\cite{zeldovich1,zeldovich2}
who proposed scattering of electromagnetic waves off a rotating cylinder,
and it has been observed in an analogue gravity experiment~\cite{Torres:2016iee}.}.
In a nutshell, low-frequency bosonic waves scattering off a rotating black hole can be superradiantly amplified 
at the expense of the black hole mass and angular momentum
if
\begin{align}
\label{eq:sprcond}
\omega_{\rm R} < & m \Omega_H
\,,
\end{align}
where $\omega_R$ is the wave's oscillation frequency,
$m$ is its spin quantum number and 
$\Omega_H$ is the angular velocity of the black hole's event horizon~\cite{Starobinsky:1973aij,Press:1973zz,Bekenstein:1973mi,
Teukolsky:1974yv}.
If, additionally, the field is confined around the black hole, 
it grows exponentially
and the system becomes unstable~\cite{Press:1972zz,Bekenstein:1998nt,Cardoso:2004nk,Brito:2015oca}.
This original gedanken experiment of a ``black hole bomb''~\cite{Press:1972zz}
can be realized in the presence of light bosons with mass $m_{\rm B} = \mu c^{-2}$
for states that satisfy $\omega_{\rm R}\lesssim\mu$~\cite{Damour:1976kh,Zouros:1979iw,Detweiler:1980uk,
Dolan:2007mj,Brito:2015oca}~\footnote{Note that asymptotically anti-de Sitter (AdS) spacetimes provide another natural scenario that yields the superradiant instability.
Although interesting in its own right, the present paper focuses on massive fields in asymptotically flat spacetimes only and refers the interested
reader to Refs.~\cite{Cardoso:2006wa,Cardoso:2013pza,Green:2015kur}.}.
Here, $\mu$ is the boson's rest-mass energy. 

The superradiant instability is strongest, i.e. growth rates are largest, if the black hole is initially highly spinning
and the gravitational coupling
$\alpha = \frac{r_{\rm g}}{\lambda_{\rm C}} =10^{10}\frac{M}{M_\odot} \frac{\mu}{\rm eV} \sim 0.42$~\cite{Dolan:2007mj,Witek:2012tr}.
To get a back-of-the-envelope estimate on the boson masses that we are sensitive to,
we rewrite the latter relation as
$\frac{\mu}{\rm eV} = 10^{-10}\,\alpha\left(\frac{M}{M_{\odot}}\right)^{-1}$.
If we consider the population of astrophysical black holes, 
i.e. the  mass range $5M_{\odot}\lesssim M \lesssim 10^{10}M_{\odot}$,
and focus on $\alpha\sim\mathcal{O}(0.1)$ 
we realize that we can probe ultralight bosons in the mass range
$10^{-12}\,{\rm eV}\gtrsim\mu\gtrsim10^{-21}\,{\rm eV}$.
This range includes the QCD axion~\cite{Peccei:1977hh},
axionlike particles proposed in the string-axiverse~\cite{Arvanitaki:2009fg},
and popular (fuzzy) dark matter candidates~\cite{Hui:2016ltb}.
That is, black holes and their observations provide a powerful tool to do 
(beyond--standard model)
particle physics in regimes that are inaccessible by traditional colliders or direct detection experiments~\cite{Arvanitaki:2010sy}.

Therefore, the superradiant instability of black holes has been studied extensively including
\begin{enumerate*}[label={(\roman*)}]
\item rigorous mathematical proofs~\cite{Shlapentokh-Rothman:2013ysa,Moschidis:2016wew};
\item perturbative calculations in the frequency~\cite{Zouros:1979iw,
Detweiler:1980uk,Dolan:2007mj,Pani:2012vp,Baryakhtar:2017ngi,
Frolov:2018ezx,Dolan:2018dqv,Siemonsen:2019ebd,Baumann:2019eav,Brito:2020lup}
and time domains~\cite{Dolan:2012yt,Witek:2012tr,East:2017mrj};
\item time evolution in the adiabatic approximation~\cite{Brito:2014wla,Ficarra:2018rfu}
and in full general relativity~\cite{Okawa:2014nda,Zilhao:2015tya,East:2017ovw};
and \item studies including axion potentials
that yield ``bosenova''-type instabilities~\cite{Yoshino:2012kn}.
\end{enumerate*}
Black hole superradiance has also been instrumental in identifying novel, hairy black hole solutions that are endowed with
complex bosons and can, thus, circumvent classical no-hair theorems~\cite{Herdeiro:2014goa,Herdeiro:2015waa,Degollado:2018ypf}.
More recently, superradiance has been explored in the context of compact binaries~\cite{Rosa:2015hoa,Blas:2016ddr,Caputo:2017zqh,Hannuksela:2018izj,Berti:2019wnn,Wong:2019kru,Bernard:2019nkv,Zhang:2019eid}.
In particular, a small object perturbing the ``gravitational atom''
yields level transitions and different types of resonances~\cite{Baumann:2019ztm}.

Most observational constraints on ultralight fields rely on a specific coupling between the QCD axion and standard model particles~\cite{Arvanitaki:2010sy,DiLuzio:2020wdo}.
Instead, the present scenario only uses minimal coupling to gravity.
Current and future gravitational wave detections have the potential
to probe for or place stringent constraints on ultralight bosons~\cite{Arvanitaki:2014wva,Hannuksela:2018izj,Cardoso:2018tly}.
In particular, 
bosons in the mass range $10^{-14}\lesssim\mu/{\rm{eV}}\lesssim10^{-11}$
are highly constrained by LIGO observations
and we expect constraints in the range $10^{-19}\lesssim\mu/{\rm{eV}}\lesssim10^{-16}$ 
from the future space-based LISA mission~\cite{Audley:2017drz,Brito:2017zvb,Isi:2018pzk,Brito:2020lup,Zhu:2020tht}.

In the present paper we focus on a different avenue and explore potentially observable signatures
of ultralight fields in the shadow of black holes.
In Refs.~\cite{Cunha:2015yba,Cunha:2019ikd}
the authors focused on the end state of the superradiant instability, i.e., after the formation of a long-lived bosonic cloud
that only slowly dissipates.
They studied, in particular, the impact of the black hole–bosonic cloud system on geodesics,
i.e., how the shadow would be affected by the changed gravitational potential.

We, instead, focus on a complementary aspect  represented by the superradiant {\textit{evolution}}. 
As the black hole undergoes the superradiant evolution its mass and angular momentum reduce as they are transferred to the bosonic condensate.
Since the shadow is determined by the black hole's parameters it may follow the superradiant evolution as well.
To quantify this statement we have combined the computation of the superradiant evolution
(in a quasiadiabatic approximation)
with that of the shadow of a Kerr black hole in a wide range of parameters.
To cleanly understand the effects of superradiant evolution, we fixed the distance of the source $r_{\rm o}$ to that of the present day value, 
and leave a more detailed analysis that takes the cosmological evolution into account for future work.
We identified two competing effects: growth of the shadow due to the reduction of the spin,
also identified in Refs.~\cite{Roy:2019esk,Davoudiasl:2019nlo},
and shrinking of the shadow due to the decrease of the black hole mass.
The shadow's angular diameter exhibits qualitatively different behavior depending on the gravitational coupling and initial black hole spin.
In particular, the black hole shadow grows due to the superradiant evolution if the coupling $\alpha$ is small,
whereas large couplings lead to a decrease of the shadow diameter.
The change of the shadow's diameter can be as large as a few $\mu$as, just below the resolution currently achievable with the EHT,
although the precise value depends on the initial black hole spin and the gravitational coupling. 

Throughout this paper, we focus on the gravitational interaction between massive scalars and rotating black holes, and its impact on the black hole shadow.
We neglect the effect of accretion by the black hole, not because its effects are small but because we wish to separate it from the signal induced by superradiance only.

\section{A brief review on the gravatom}\label{sec:Gravatom}
\subsection{Setting the stage}
The ``gravatom,'' or ``gravitational atom,'' refers to a Kerr black hole surrounded by a long-lived cloud composed of ultralight bosonic fields. This cloud develops as a consequence of the superradiant instability: low-frequency bosonic fields that meet condition~\eqref{eq:sprcond} are superradiantly amplified at the expense of the black hole's mass and angular momentum. These fields are massive with mass $m_{\rm B}= \mu c^{-2}$ and they can be trapped in the vicinity of the black hole if $\omega_{\rm R}\lesssim \mu$ (see below), and grow exponentially until the superradiant condition~\eqref{eq:sprcond}
is saturated.

Here, we focus on massive scalar fields $\Phi\sim e^{-\imath\omega t} R(r) Y_{lm}(\theta,\varphi)$
that satisfy the Klein-Gordon equation in a Kerr background,
and give a brief review of its key features. 
The precise superradiant evolution crucially depends on the dimensionless gravitational coupling
\begin{align}
\label{eq:DefGravCoupling}
\alpha = & \frac{G\,M}{c^2} \frac{\mu}{c\,\hbar}
       =   \frac{r_{\rm g}}{\lambda_{\rm C}}
       \simeq 10^{10} \left(\frac{M}{M_{\odot}}\right) \left(\frac{\mu}{\rm eV}\right)
\,,
\end{align}
where $M$ is the black hole mass and $\mu=m_{\rm B} c^2$ the scalar field's rest-mass energy. 
The gravitational coupling is determined by the ratio between the black hole's gravitational radius
$r_{\rm g}=\frac{GM}{c^{2}}$ and the field's reduced Compton wavelength $\lambda_{\rm c}=\frac{\hbar}{m_{\rm B}\,c}=\frac{\hbar\,c}{\mu}$. 
In the following we will use Planck units $G=1$, $c=1$, $\hbar=1$.

To explore the phenomenology of massive scalars around rotating black holes one solves
the Klein-Gordon equation in the Kerr spacetime~\cite{Dolan:2007mj}.
In the limit $\alpha\ll1$ this calculation simplifies to a Schr\"{o}dinger-type equation 
that we can solve analytically for the scalar's complex 
(or {\textit{quasinormal}} mode) 
frequency $\omega = \omega_{\rm R} + \imath \Gamma$~\cite{Detweiler:1980uk}.
The spectrum of oscillation frequencies for modes $(nlm)$ is
\begin{align}
\label{eq:SFFreqRe}
\omega_{{\rm R},nlm} = & \mu\left[1-\frac{1}{2}\left(\frac{\alpha}{n+l+1}\right)^2\right]\sim\mu
\,,
\end{align}
where $n=0,1,\ldots$, $l=0,1,\ldots$ and $-l\leq m \leq +l$ are the scalar's
principal, azimuthal and spin quantum numbers.
Note that it resembles the spectrum of a hydrogen atom and, thus, inspired the terminology ``gravatom''.
In contrast to the hydrogen atom, however, 
the frequency is complex and its imaginary part determines 
the scalar's growth ($\Gamma>0$) or decay ($\Gamma<0$) 
rate on $e$-folding timescales
$\tau_{\rm SR}=1/|\Gamma|$. For each mode $(nlm)$, it is given by
\begin{align}
\label{eq:SFFreqIm}
\Gamma_{nlm} = & -2\frac{r_{+}}{r_{\rm g}}(\mu-m\Omega_H)\alpha^{4l+5}\sigma_{nlm}
\,,
\end{align}
where 
$\Omega_{\rm H}=\frac{\chi}{2r_{+}}$ 
is the angular velocity of the black hole's horizon
at radius $r_{+}=r_{\rm g}(1+\sqrt{1 - \chi^{2}})$,
$\chi\equiv{J/M^2}$ the dimensionless spin, 
and
\begin{align}
\sigma_{nlm} \equiv & \frac{2^{4l+1}(2l+n+1)!}{(l+n+1)^{2l+4}(n!)}\Big[\frac{l!}{(2l)!(2l+1)!}\Big]^2
\nonumber\\ \times &
\prod_{k=1}^{l}\Big[k^2\Big(1-\chi^2\Big)+4r_{+}^2(\mu-m\Omega_H)^2\Big]
\,.
\end{align}
If $\omega_{\rm R} \sim \mu < m\Omega_{\rm H}$~\footnote{In the following we suppress subscripts $(nlm)$ unless they are needed explicitly.}
the imaginary part of the frequency becomes positive, see Eq.~\eqref{eq:SFFreqIm}, 
and the scalar fields grow exponentially.
Equation~\eqref{eq:SFFreqIm} furthermore implies that the fastest growing mode corresponds to the lowest-lying value of the orbital quantum number $l$ and $m=l>0$. 
For massive scalar fields this is the dipole mode $l=m=1$. 

We remark that only sufficiently rapidly rotating black holes develop the superradiant instability.
Combining the superradiant condition~\eqref{eq:sprcond}
and the relation $\omega_{\rm R}\simeq\mu$, Eq.~\eqref{eq:SFFreqRe},
we find that only black holes with a dimensionless spin
\begin{align}
\label{eq:sprcondSpin}
\chi \geq & \chi_{\rm crit} = \frac{4m\alpha}{m^2 + 4 \alpha^2}
\,,
\end{align}
undergo superradiant scattering.
Note, that the derivation assumed $M\omega_{\rm R}\sim M\mu=\alpha\neq0$ so the limit $\alpha\to0$ is not well-defined.

\subsection{Quasiadiabatic evolution}\label{sec:Evolution}
We are interested in the 
effect of the superradiant evolution on the time development of the black hole shadow 
which, in turn, is determined by the evolution of the hole's mass and spin. 
The key phases of the superradiant evolution are
\begin{description}[noitemsep,leftmargin=*,labelindent=0mm]
\item [0] Superradiant evolution starts for a black hole of 
initial mass $\MZ$ and initial spin $\chiZ$, 
surrounded by a scalar condensate of initial total mass $\McZ$.
The scalar field condensate grows at the expense of the black hole's energy and angular momentum.
\item [I] The superradiant evolution continues until condition~\eqref{eq:sprcond} is saturated. 
We refer to phase~I as the phase in which the black hole parameters have the largest gradients in time.
te of the gravatom, during which the black hole of (final) mass 
$\MII$ and spin $\chiII$
is surrounded by a long-lived scalar condensate that now has a total mass $\McII$~\footnote{Here, we focus on real scalar fields that slowly decay over time. 
Complex scalars can give rise to nonlinear, hairy black hole solutions that appear at the onset of the 
superradiant instability~\cite{Herdeiro:2014goa,Herdeiro:2015waa}.
};
\item [III] Dissipation of the bosonic cloud due to gravitational wave emission.
\end{description}
We define the onset of phase~I as the time $\tI$ where the scalar cloud has acquired a mass 
$M_{c}(\tI) = \McZ + \Z \MZ$.
The coefficient $\Z$ has an appropriately chosen, fixed value. In particular, 
we specify $\Z=10^{-5}$ ($\Z=10^{-4}$) for small (large) scalar cloud seeds.
We give an explicit expression for $\tI$ in  Sec.~\ref{ssec:ReachingPhaseI}, 
based on the analytic approximation presented in Sec.~\ref{sec:AnalyticApproximation}. 

Following Refs.~\cite{Brito:2014wla,Ficarra:2018rfu},
we work in the small coupling regime, $\alpha\ll1$, and evolve the black hole--cloud system in a quasiadiabatic approximation.
We are interested in the evolution of the black hole's parameters that have a direct impact on the shadow and that can change during the superradiant buildup of the bosonic condensate, i.e. phases 0 to II.
The black hole remains unaffected by the dissipation of the cloud in phase III~\cite{Brito:2014wla,Ficarra:2018rfu}.
Furthermore, the cloud decays due to gravitational wave emission on timescales $\tau_{\rm GW}$ 
much larger than the superradiance timescale $\tau_{\rm SR}=\frac{1}{\Gamma}$~\cite{Yoshino:2013ofa,Baumann:2018vus}.

Before we review the quasiadiabatic evolution, let us quantify the latter statement.
The gravitational wave energy flux can be approximated by~\cite{Yoshino:2013ofa}
\begin{align}
\label{eq:CloudGWEmission}
\frac{\dif E_{\rm GW}}{\dif t} = & C_{nl} \left(\frac{M_c}{M}\right)^2 \alpha^{4l+10}
\,,
\end{align}
where $M_{c}$ and $M$ are, respectively, the cloud and the black hole mass,
$\alpha=M\mu$ the gravitational coupling, and~\footnote{Note, that the authors of Ref.~\cite{Yoshino:2013ofa} use $n=l+1+n_{r}$, $n_{r}=0,1,\ldots$. 
In our convention $n=n_{r}=0,1,\ldots$ and we have redefined the coefficient accordingly.}
\begin{align}
\label{eq:CloudGWEmissionCoefficient}
C_{nl} = & \frac{ 16^{l+1} l (2l-1) \Gamma[2l-1]^2 \Gamma[ n+2(l+1) ]^2 }{ (n+l+1)^{4l+8} (l+1) \Gamma[l+1]^4 \Gamma[4l+3] \Gamma[n+1]^2 }
\,.
\end{align}
At this stage, the system's evolution is dominated by transforming energy from the cloud into gravitational radiation
that leaves the black hole parameters essentially unchanged.
We can then use the energy conservation relation
\begin{align*}
\frac{\dif E_{\rm GW}}{\dif t} = & - \frac{\dif M_{\rm c}}{\dif t}
\end{align*}
to integrate Eq.~\eqref{eq:CloudGWEmission} to
\begin{align}
\label{eq:CloudMassGWEmission}
M_{\rm c}(t) = & \McII \left[ 1 + \frac{t}{\tau_{\rm GW}} \right]^{-1}
\,,
\end{align}
where $\McII$ is the mass of the cloud in phase~II, i.e., after the superradiant evolution but before gravitational wave emission becomes effective. We introduced the gravitational wave emission timescale $\tau_{\rm GW}$
\begin{align}
\label{eq:TauGW}
\tau_{\rm GW} = & \frac{m\,M_{0}}{\chi_{0}\,C_{nl}} \alpha^{-(4l+11)}
\\ = &
        1.25\times10^5 \rm yr
       \left(\frac{M_0}{M_{\odot}}\right) \left(\frac{\chi_0}{0.8}\right)^{-1} \left(\frac{\alpha}{0.1}\right)^{-15}
\,,\nonumber
\end{align}
where we used the approximation
$\McII \sim \frac{\omega_{R}}{m} J_{\rm 0} \sim \frac{\alpha\,\MZ \chiZ}{m}$~\cite{Brito:2017zvb},
and in the second line we provide an estimate for the dominant mode $l=m=1$.
This result is in good agreement with numerical computations of the superradiant evolution that include gravitational wave emission~\cite{Brito:2014wla,Ficarra:2018rfu}.
Finally, the ratio of the gravitational wave timescale $\tau_{\rm GW}$ 
and the superradiance timescale $\tau_{\rm SR}$ is 
\begin{align}
\frac{\tau_{\rm GW}}{\tau_{\rm SR}} = & 
- \frac{2 m r_{+}}{\chi_0} \frac{\sigma_{nlm}}{C_{nl}} \left(\mu-m\Omega_{\rm H} \right) \alpha^{-6}
\,.
\end{align}
For a dipole ($l=m=1$) scalar cloud with coupling $\alpha=0.1$ around a black hole with initial
dimensionless spin $\chiZ=0.8$, this ratio is $\frac{\tau_{\rm GW}}{\tau_{\rm SR}} \sim  10^7$. 
It is comparable for other spin values and 
increases as we decrease the gravitational coupling.
As the order of magnitude indicates, the depletion of the cloud is not significant in the evolution of the shadow until very late times. Hence, we conclude that the gravitational wave emission period does not interfere with the superradiance evolution, and so we can safely ignore this effect for present purposes.

If, additionally, we neglect external processes such as accretion of ordinary matter
whose effects have been found to be subdominant~\cite{Brito:2014wla},
we can describe the superradiant evolution by 
\begin{subequations}
\label{eq:spinevolutionsdiffeqs}
\begin{align}
\frac{dJ}{dt}   = & -\frac{dJ_c}{dt}
\,,\\
\frac{dM}{dt}   = & -\frac{dM_c}{dt}
\,,\\
\frac{dJ_c}{dt} = & \frac{m}{\mu}\frac{dM_c}{dt}
\,.
\end{align}
\end{subequations}
The first two equations correspond to energy and momentum conservation, whereas the last one can be understood 
as a balance equation between angular momentum (``quanta'' with momentum $\hbar\,m$) and energy (quanta with energy $\hbar\omega\sim\hbar\mu$) due to perturbations. A detailed derivation can be found in Ref.~\cite{Brito:2015oca}.
The energy of the cloud 
evolves as \cite{Brito:2014wla}
\begin{align}
\label{eq:energyinstabilityflux}
\frac{dM_c}{dt} = & 2\Gamma_{nlm}M_c
\,.
\end{align}
Now we can solve the system of differential equations numerically, given suitable initial data.
While our computations are valid for a wide range of black hole parameters,
we exemplarily present our results for a M87-like black hole and set the initial mass to 
$M_{0}=M_{\rm M87} = 6.5\times 10^{9}M_{\odot}$.
For the scalar field we choose either a small seed $\McZ=10^{-9}M_{0}$ that mimics small (``quantum'') fluctuations
or a large seed of $M_{c,0}=0.025 M_{0}$ that may be present after the merger of two gravitational atoms or 
toward the end of the superradiant evolution~\cite{Brito:2014wla,Ficarra:2018rfu,Okawa:2014nda}.
We consider an ultralight scalar with mass 
$\mu=1\times{10^{-21}}~\text{eV}$,
so the gravitational coupling is $\alpha \sim 0.05$.
We have simulated the superradiant evolution for different values of the initial spin $\chi_{0}$ and the dominant
superradiant mode of the scalar, i.e., $l=m=1$.
\begin{figure*}[htpb!]
\centering
\includegraphics[width=0.94\textwidth,clip]{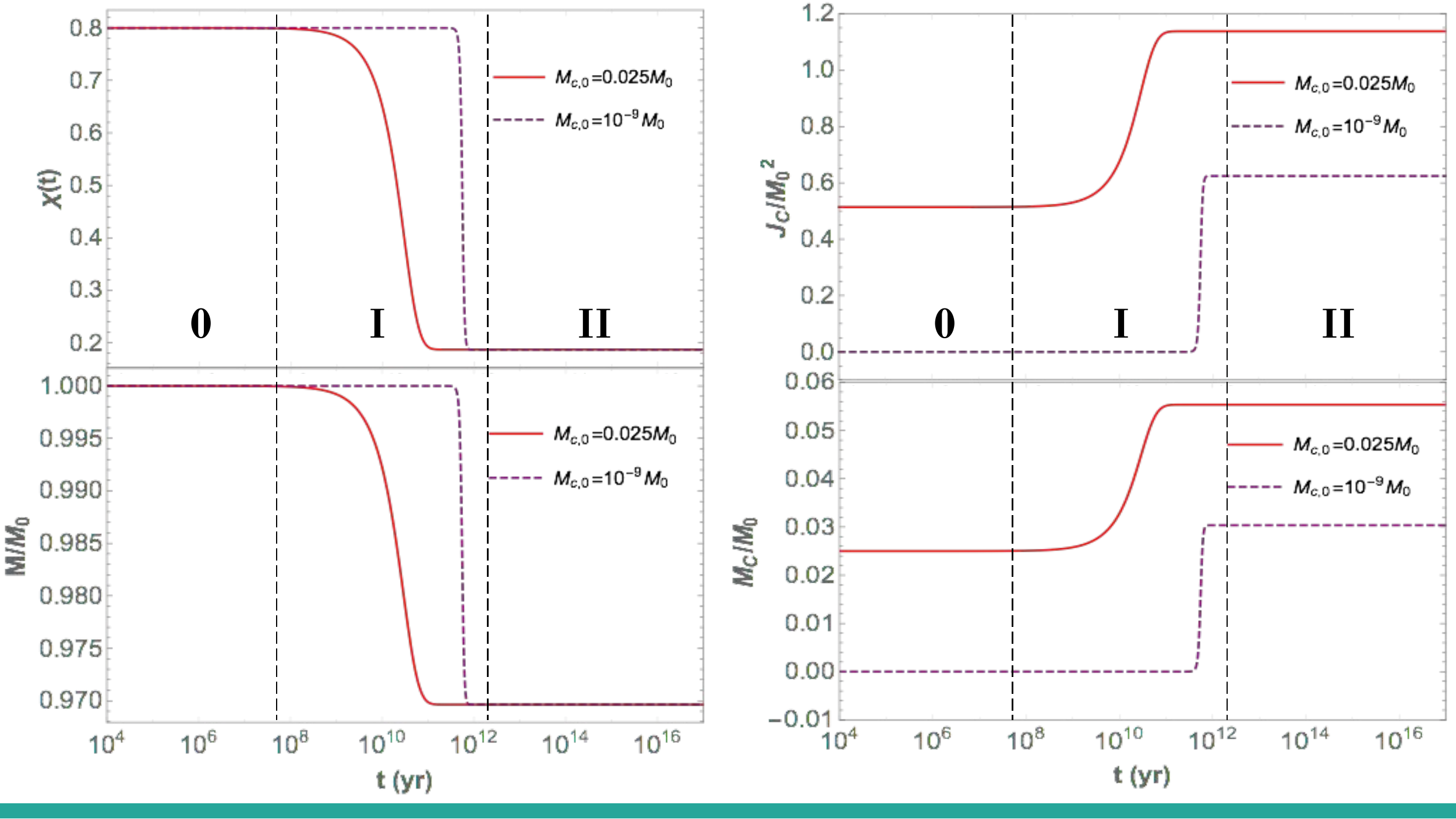}
\caption{\label{fig:Spinevolutionseeds}
Superradiant evolution of an ultralight scalar with $\mu=10^{-21}$eV 
surrounding a black hole with initial mass $M_{0}=M_{\rm M87}=6.5\times10^{9}M_{\odot}$.
We consider both a small and a large scalar seed, i.e., $M_{c,0}=10^{-9}M_{0}$ (purple, dashed curves)
and $M_{c,0}=0.025M_{0}$ (red, solid curves).
We indicate, qualitatively, 
phases~0--II of the superradiant evolution.
Left: Evolution of the black hole's spin (top) and mass (bottom).
Right: Evolution of the scalar cloud angular momentum (top) and cloud mass (bottom) normalized by the initial black hole mass.}
\end{figure*}
Exemplarily, 
we present the evolution for a system with $\chi_{0}=0.8$ and both types of seeds in Fig.~\ref{fig:Spinevolutionseeds}.
This enables us to verify our simulations against results available in the literature~\cite{Brito:2014wla,Ficarra:2018rfu}.

Additionally, it is useful to derive exact analytic expressions for the black hole's parameters 
at the end of the superradiant evolution indicated by phase~II. This is signaled by the saturation
of the superradiant condition~\eqref{eq:sprcond}.
At this stage, the final black hole spin  $\chiII=\chi_{\rm crit}$ 
with the critical spin given in Eq.~\eqref{eq:sprcondSpin}. 
The change of the black hole mass $M$ and spin $J=\chi\,M^2$ are related via
\begin{align}
\label{eq:finalmassderiv}
\MII-\MZ = & \frac{\mu}{m} \left(J_{\rm II} - J_{\rm 0} \right)
        = \frac{4 \alpha_{\rm II}^2 \MII}{m^2+4\alpha_{\rm II}^2} - \frac{\alpha\chiZ\MZ}{m} 
\,,
\end{align}
where $\alpha_{\rm II} = \MII\mu$ and we used Eq.~\eqref{eq:sprcondSpin}.
Note that $\frac{\alpha_{\rm II}}{\alpha} = \frac{\MII}{\MZ}$, and we denoted here $\alpha=\alpha_0$ to avoid too many subscripts. In fact we use throughout the text the notation that $\alpha$ denotes the initial coupling, unless stated otherwise or clear from the context.
Solving the polynomial for $\MII$ yields
\begin{eqnarray}\label{eq:finalmassBH}
\frac{\MII}{\MZ} &=& \frac{m^3}{8\alpha^2 \left(m-\alpha\chi_{0}\right) }
\left\{ 1 - \sqrt{1 - \frac{16\alpha^2 \left(m - \alpha\chi_{0} \right)^2}{m^4} } \right\}
\nonumber\\
&\simeq&1-\alpha\frac{\chi_0}{m}+{\cal O}(\alpha^2)\ ,
\end{eqnarray}
where in the last expression we imposed the small coupling approximation. 
Using this, we can rewrite the formula for the final spin in terms of the initial spin and coupling. One finds
\begin{equation}
\label{eq:finalBHspin}
\chi_{\rm II}=\frac{4m\Big(\frac{M_{\rm II}}{M_0}\Big)\alpha}{m^2+4\Big(\frac{M_{\rm II}}{M_0}\Big)^2\alpha^2}\ ,
\end{equation}
where one plugs in \eqref{eq:finalmassBH} as a function of $\chi_0$ and $\alpha$.
It is  then an easy exercise to show that $\chi_{\rm II}\geq \chi_0$ as it should indeed.
We find excellent agreement of these analytical formulas with the numerical result with $\lesssim0.0015\%$ relative error.

\section{Shadow and superradiance}\label{sec:shadowsuperradiance}
\subsection{Review and definitions}
Gravitational lensing is the phenomenon by which light gets deflected due to the gravitational influence of a massive body. 
Such deflection of electromagnetic radiation, e.g. emitted by the accretion disk surrounding a black hole,
gives rise to what we call the black hole ``shadow.''

The strategy for studying gravitational lensing consists of computing the geodesics of photons reaching the observer.
The specific shape of the observed shadow depends on a variety of parameters such as the orientation to the observer 
and the black hole's mass and spin. 
We denote the orientation of an observer facing the 
equatorial plane as $\theta_{o}=0$ (face-on)
and the orientation of an observer lying in the equatorial plane as $\theta_{o}=\pi/2$ (edge-on); see illustration in Fig.~\ref{fig:ObsSky}.

\begin{figure}[h!]
\begin{center}
\includegraphics[width=0.49\textwidth,clip]{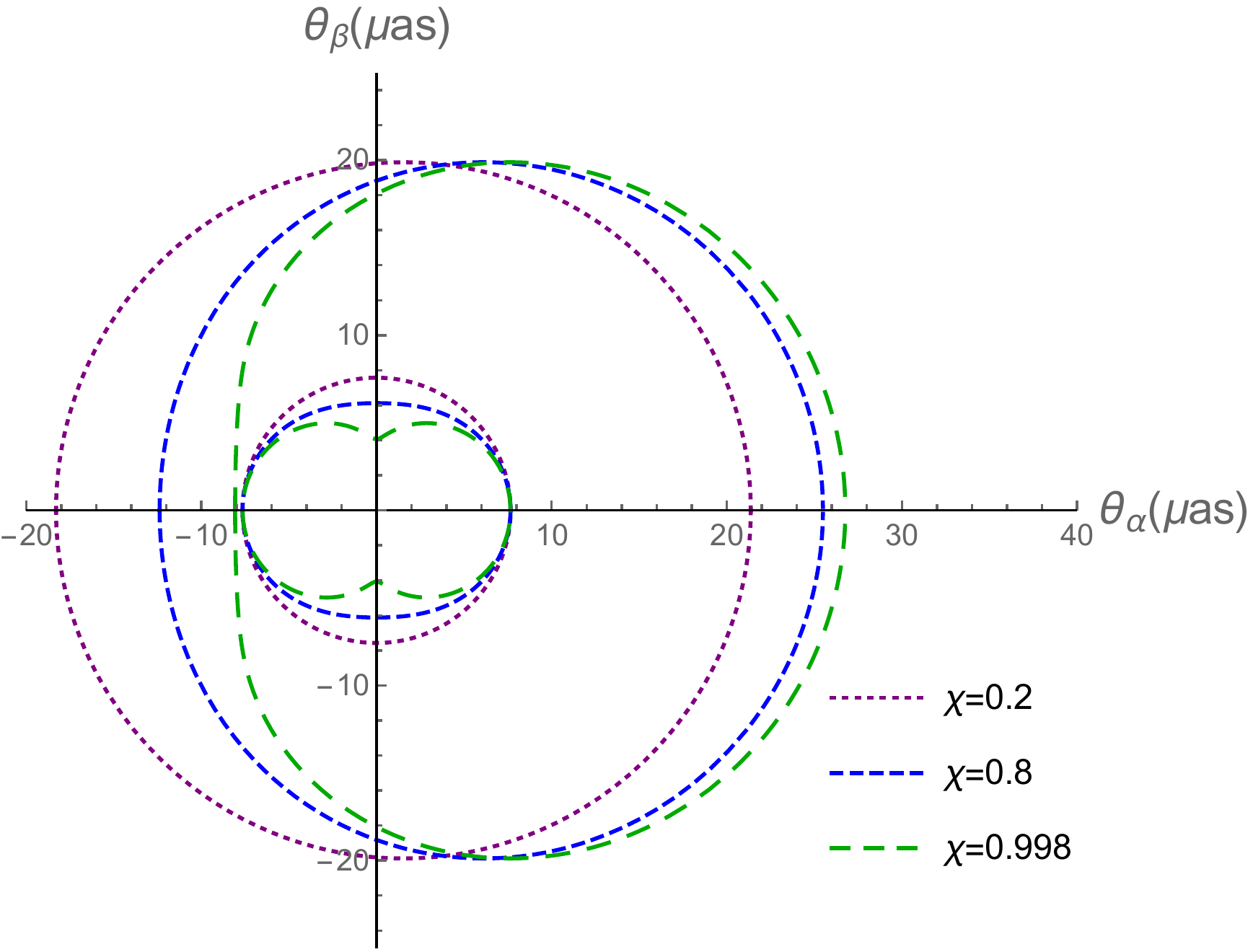}
\caption{\label{fig:KerrBHShadow}
Inner edge of the shadow $d_{\rm sh}$ and the outer horizon $r_{+}$ of a Kerr black hole with dimensionless
spins $\chi=0.2$ (purple dotted curve), $\chi=0.8$ (blue dashed curve) and $\chi = 0.998$ (green long-dashed curve)
as seen edge-on.
Exemplarily, we set the mass $M=6.5\times10^{9}M_{\odot}$ and distance $r_{o}=16.8\text{Mpc}$, i.e., parameters corresponding to M87.}
\end{center}
\end{figure}

For rapidly rotating black holes prograde photons are deflected closer to the black hole horizon than retrograde photons and yield a highly asymmetric shadow, whereas slowly rotating black holes exhibit a (nearly) spherical shadow as illustrated in Fig.~\ref{fig:KerrBHShadow}.
To derive the location $\bar{r}$ of the inner edge
(or ``maximum approach distance'')
of the black hole shadow we have to solve the constants of motion in a Kerr spacetime, together with the equation for the Carter constant.
We provide the detailed derivation in Appendix~\ref{app:shadowkerrbh} and here only present the results. 
For simplicity, we focus on an observer oriented edge-on.
The shadow is determined by the parametric curve $(\theta_\alpha(\bar{r}),\theta_\beta(\bar{r}))$,
where $\theta_{\alpha/\beta}(\bar{r})$ is the angular separation in observer sky coordinates
as depicted in Fig.~\ref{fig:KerrBHShadow}.
As a measure for the shadow size and because it changes most significantly along $\theta_{\beta}=0$, we define its (angular) diameter as
\begin{align}
d_{sh}\equiv|\theta_{\alpha}(\bar{r}_{+})|+|\theta_{\alpha}(\bar{r}_{-})| ~,
\end{align}
with $\bar{r}_{+}$ ($\bar{r}_{-}$) being the maximum approach distance for prograde (retrogade) photons. 
Since $\bar{r}_{\pm}$ cannot be expressed in a simple way (at least for arbitrary values of the spin, mass and observer inclination) we have computed it numerically. In order to gain a better understanding of how the shadow changes with spin and orientations we will fix one of them and vary the other. 

\begin{figure}[htpb!]
\begin{center}
\includegraphics[width=0.48\textwidth,clip]{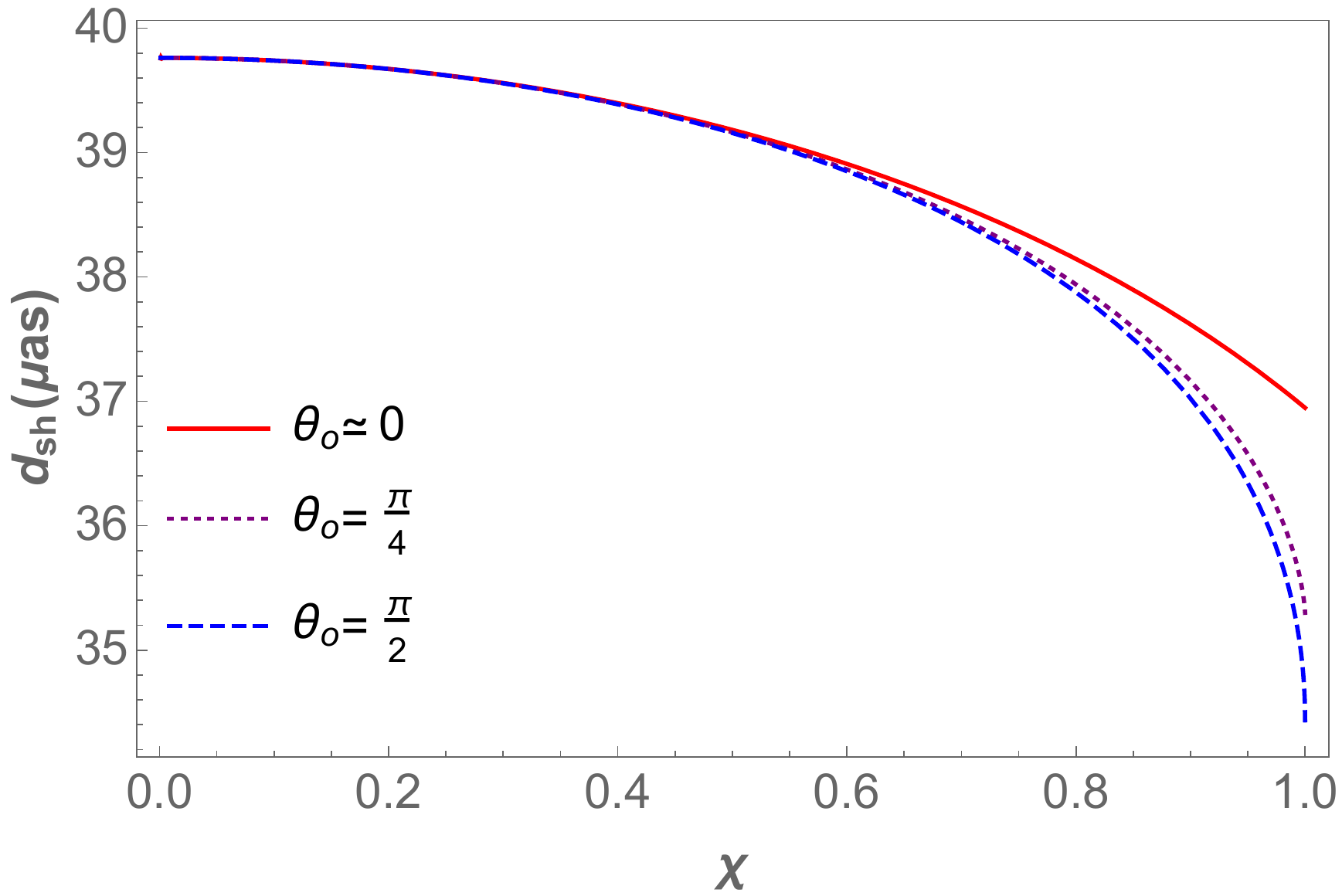}
\caption{\label{fig:AngulardiameterBHshadowsh}
Angular diameter $d_{\rm sh}$ as a function of the (dimensionless) black hole spin $\chi$ for different
observer orientations $\theta_{o}$.
Due to axial symmetry, all orientations $\theta_{o}>\pi/2$ can be recovered from the interval $0\leq\theta_{o}\leq\pi/2$.
Exemplarily, we set $M=6.5\times10^{9}M_{\odot}$ and $r_{o}=16.8$Mpc. 
}
\end{center}
\end{figure}

\noindent{\textbf{Fixing the orientation:}}
As illustrated in Fig.~\ref{fig:KerrBHShadow}, the shadow diameter $d_{\rm sh}$ depends on the spin of the central black hole.
We show this dependence in more detail in Fig.~\ref{fig:AngulardiameterBHshadowsh} where we present the shadow diameter, 
exemplarily for a M87-like black hole with mass $M = 6.5\times10^9M_{\odot}$,
as a function of the dimensionless spin and for different fixed orientations $\theta_{o}$ of the observer.
We see that, due to the axial symmetry, the diameter seen by an observer facing the pole $\theta_o=0$ (i.e., face-on) 
changes less than the diameter seen by an equatorial observer (i.e., edge-on), where the deformation is maximal. 
Apart from $\theta_o=0$, intermediate orientations are close to the equatorial case.

\begin{figure}[htpb!]
\begin{center}
\includegraphics[width=0.48\textwidth,clip]{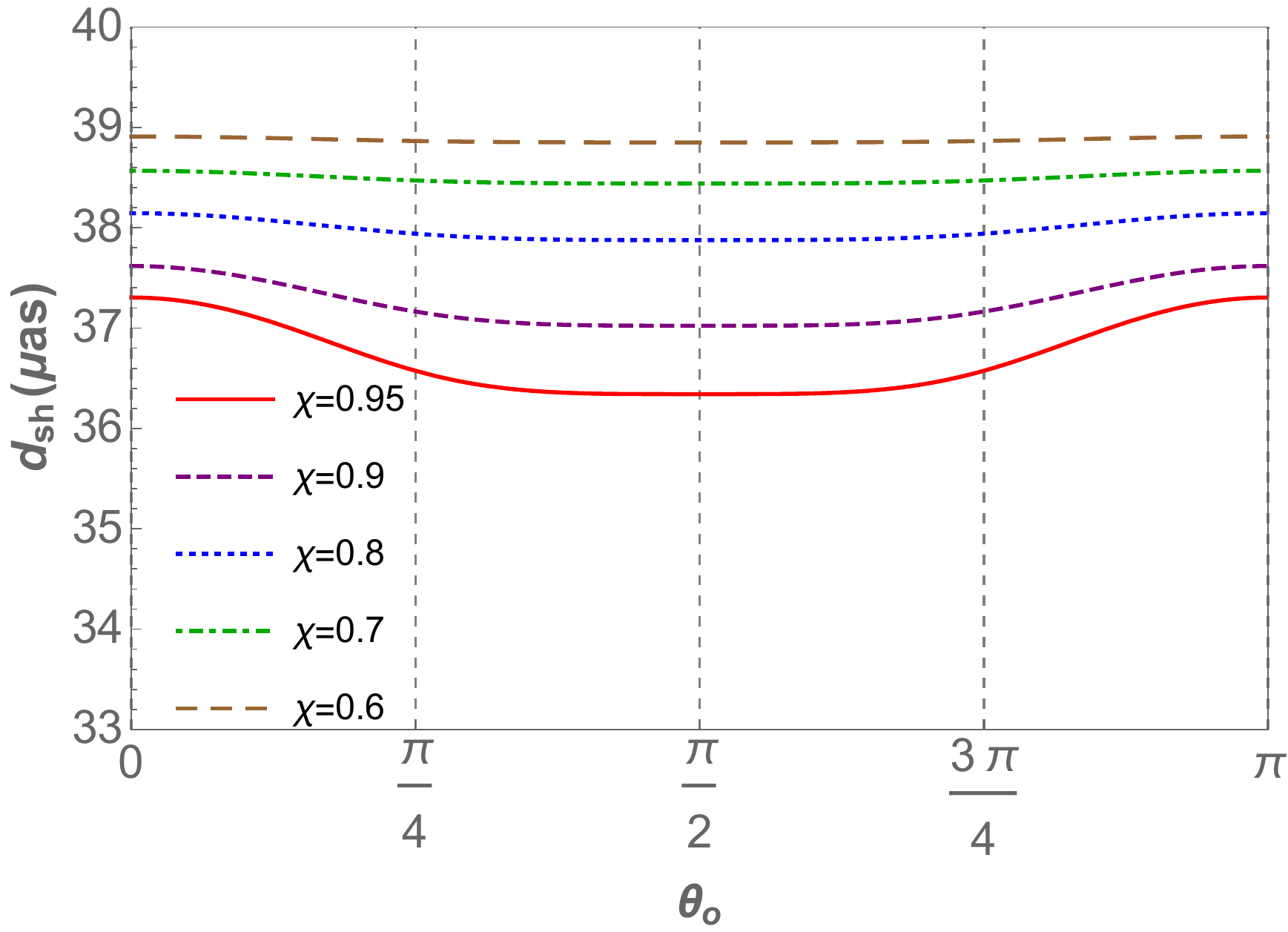}
\caption{\label{fig:AngulardiameterBHshadowshorientations}
        Shadow diameter $d_{\rm sh}$ as function of the orientation $\theta_{o}$ toward an observer for
        different values of the dimensionless black hole spin $\chi$.
}
\end{center}
\end{figure}

\noindent{\textbf{Fixing the spin:}}
We illustrate the dependence of the angular diameter $d_{\rm sh}$ as a function of the observer's orientation $\theta_{o}$ for various fixed values of the spin in Fig.~\ref{fig:AngulardiameterBHshadowshorientations}.
We observe that the shadow diameter depends more strongly on the observer's angle as the black hole spin increases. This is not surprising since small spins yield almost spherically shaped shadows whereas high spins lead to asymmetric shapes.
Furthermore, the angular diameter of a shadow is degenerate for different values of the black hole spin and orientation. That is, even after fixing the black hole's mass and distance, the same $d_{\rm sh}$ could correspond to different pairs $(\chi,\theta_{o})$. To break this degeneracy, we need independent measurements of the orientation or of the black hole spin, e.g., using the methods reviewed in~\cite{McClintock:2011zq,Garcia:2019nac,Reynolds:2019uxi}.

\subsection{Shadow time evolution}\label{ssec:ShadowEvolution}
\begin{figure*}[htpb!]
\begin{center}
\subfloat[$\McZ=10^{-9}\MZ$]{\includegraphics[height=0.25\textheight,clip]{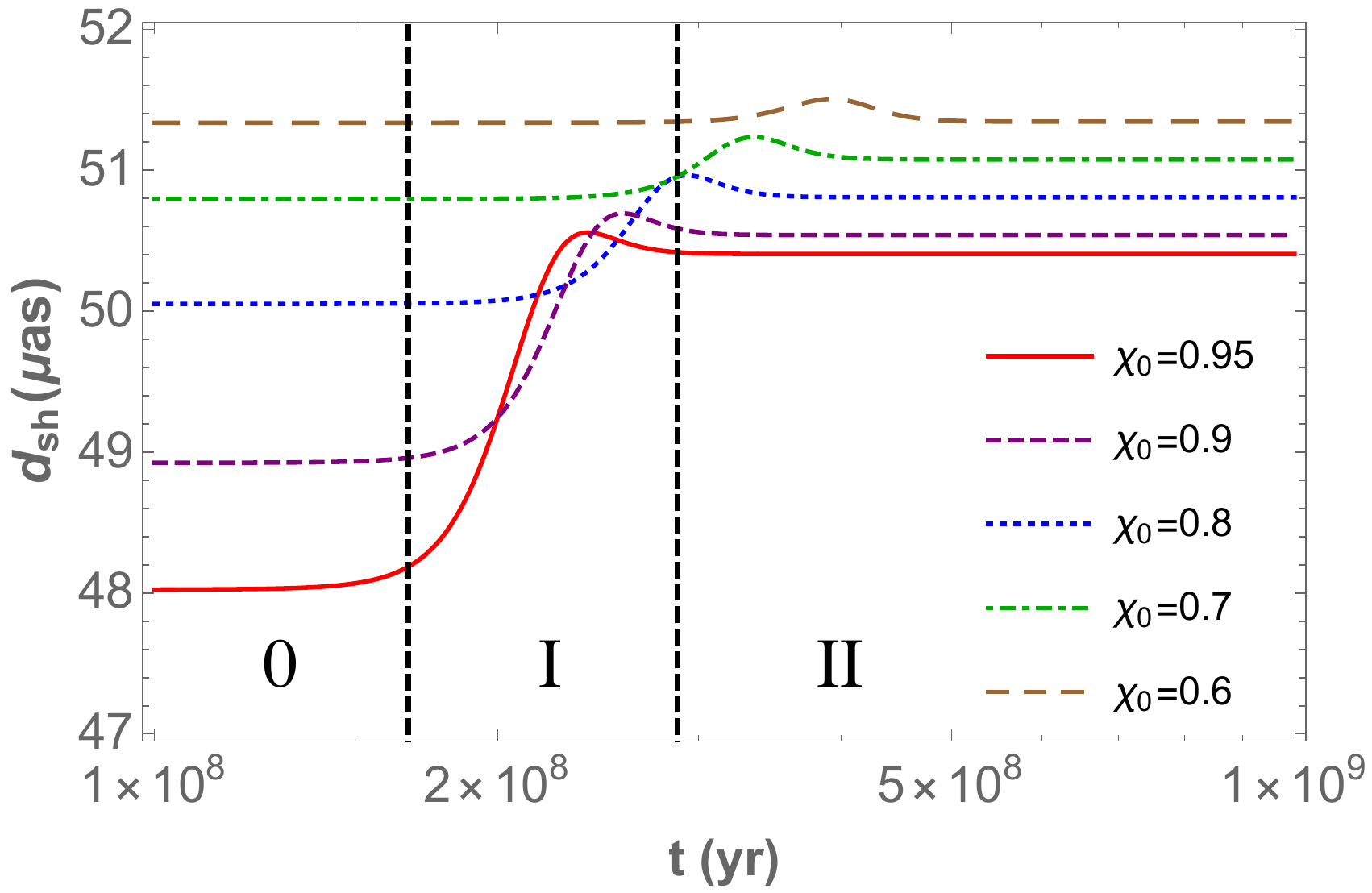}\label{fig:AngulardiameterevolutionSgrASmallSeed}}
\subfloat[$\McZ=0.025\MZ$  ]{\includegraphics[height=0.25\textheight,clip]{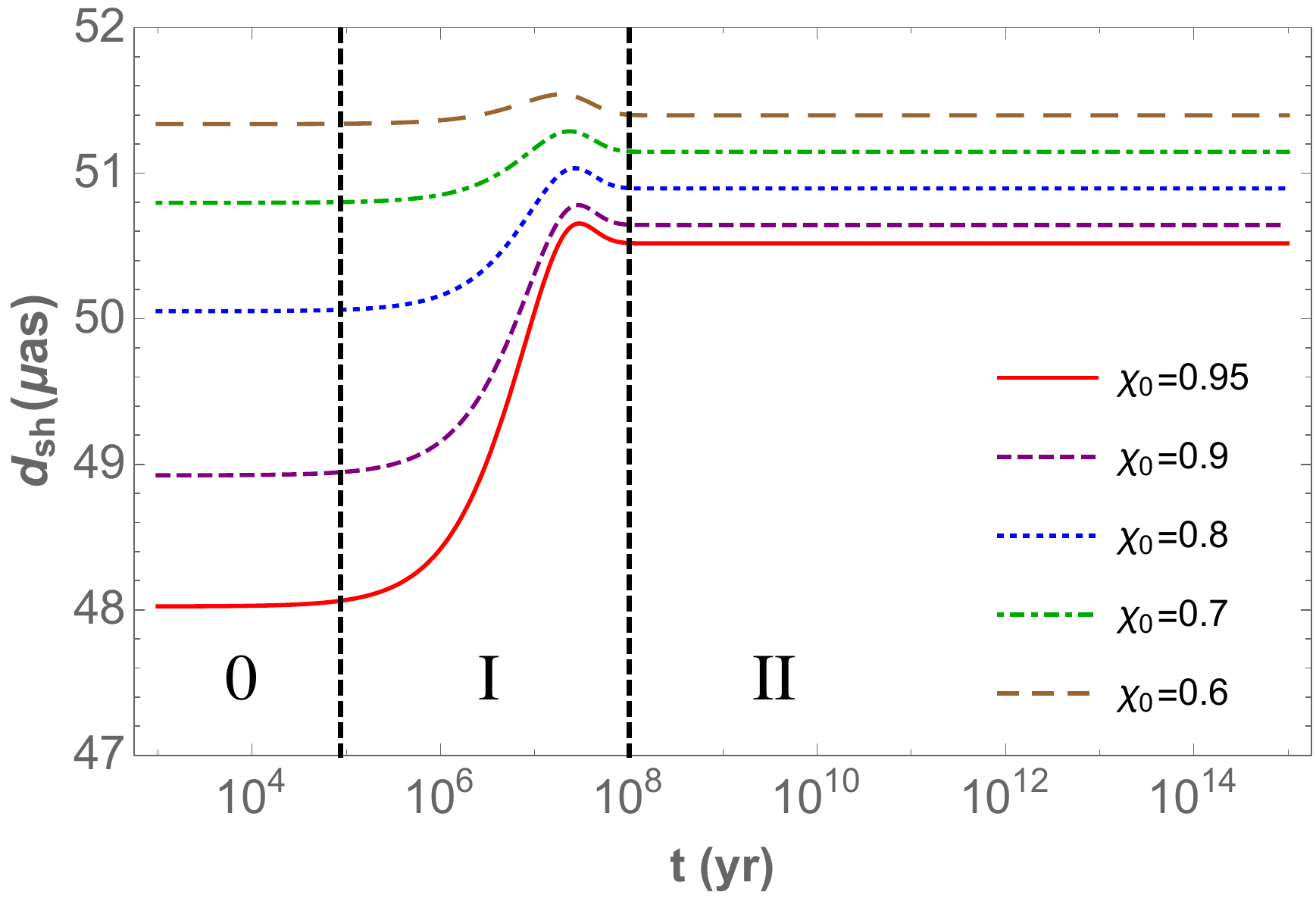}\label{fig:AngulardiameterevolutionSgrALargeSeed}}
\caption{\label{fig:AngulardiameterevolutionSgrA}
        Evolution of the shadow diameter $d_{\rm sh}$ for a Sgr~A$^{\ast}$-type black hole with initial mass $\MZ=4.2\times10^6M_{\odot}$ 
        and different initial spins $\chiZ$, at a distance of $r_{\rm o}=8.2$kpc.
        The gravitational coupling is  $\alpha=0.05$ so that the scalar condensate is composed of particles with mass 
        $\mu=1.5\times10^{-18}$eV. The initial cloud mass is $\McZ=10^{-9}\MZ$ (left) or $\McZ=0.025\MZ$ (right). 
        Here we show the different stages (see Sec.~\ref{sec:Gravatom}) exemplarily for the case of $\chiZ=0.95$.
}
\end{center}
\end{figure*}

\begin{figure*}[htpb!]
\begin{center}
\subfloat[$\McZ=10^{-9}\MZ$]{\includegraphics[height=0.24\textheight,clip]{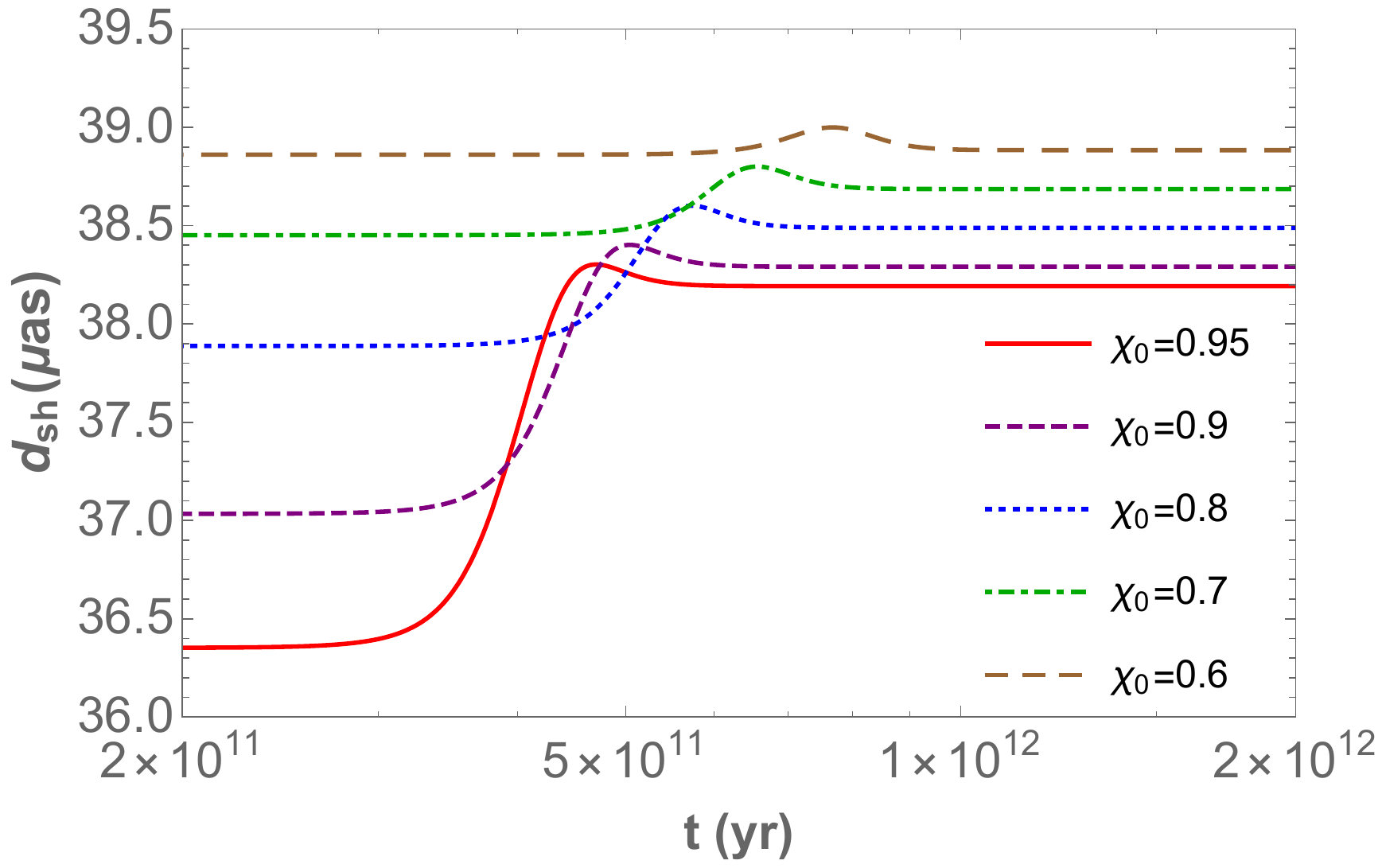}\label{fig:AngulardiameterevolutionM87SmallSeed}}
\subfloat[$\McZ=0.025\MZ$  ]{\includegraphics[height=0.24\textheight,clip]{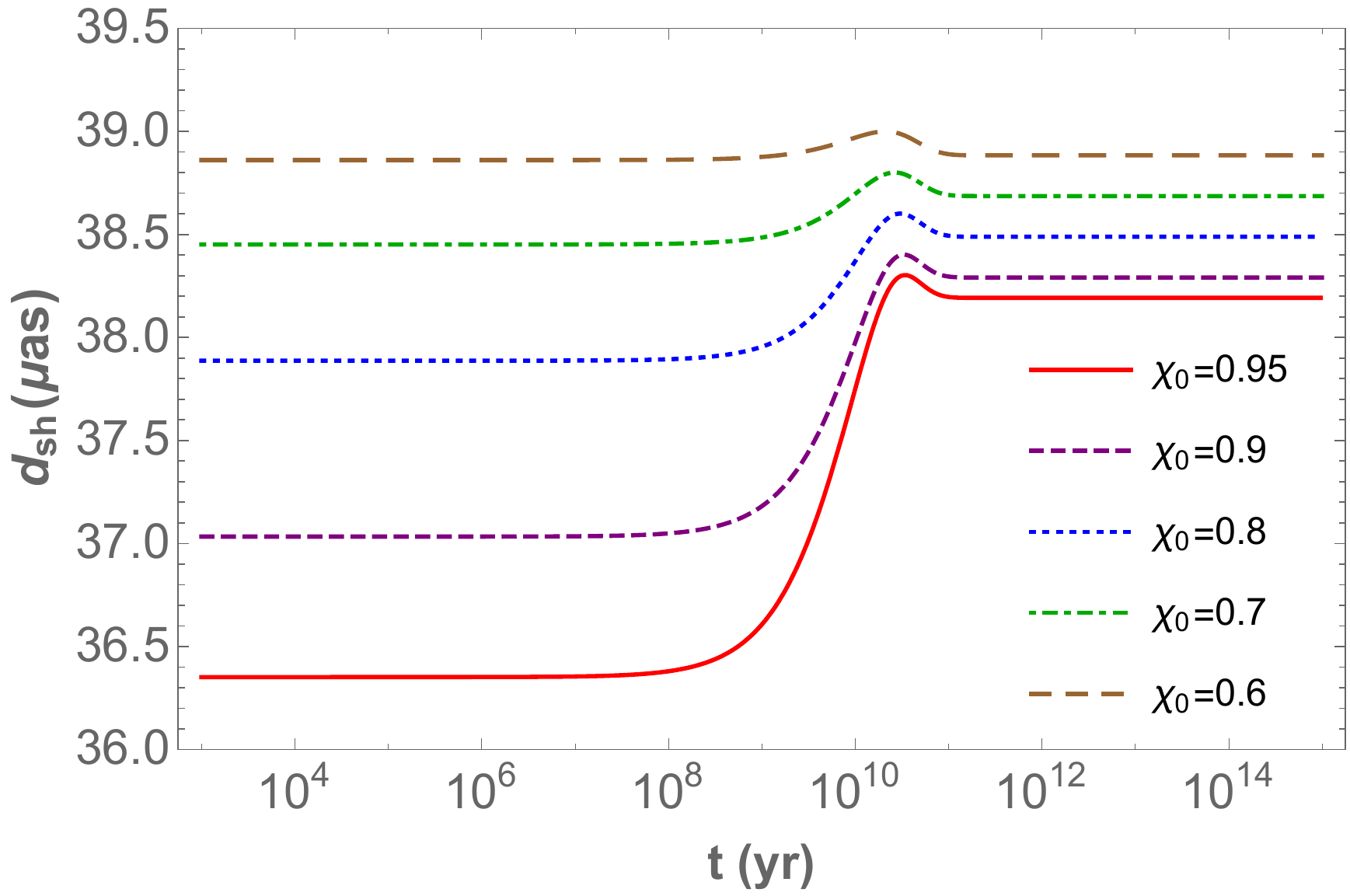}\label{fig:AngulardiameterevolutionM87LargeSeed}}
\caption{\label{fig:AngulardiameterevolutionM87}
        Evolution of the shadow diameter $d_{\rm sh}$ for a M87-type black hole with initial mass $\MZ=6.5\times10^9M_{\odot}$ 
        and different initial spins $\chiZ$, at a distance of $r_{\rm o}=16.8$Mpc.
        The gravitational coupling is $\alpha=0.05$, i.e., the scalar condensate is composed of particles with mass 
        $\mu=10^{-21}$eV. The initial cloud mass is $\McZ=10^{-9}\MZ$ (left) or $\McZ=0.025\MZ$ (right).
}
\end{center}
\end{figure*}

As we have seen, the shadow's shape and angular diameter depend on the mass and spin of the central black hole.
Here, we explore how the superradiant evolution 
affects the black hole shadow. 
There are two competing influences at play:
the decrease of the mass of the black hole would lead to a {\textit{decrease}} of the angular diameter, i.e., $\dif d_{\rm sh} / \dif M > 0$;
the decrease of the black hole spin would lead to an {\textit{increase}} of the angular diameter, i.e., $\dif d_{\rm sh}/\dif J < 0$.
To understand this behavior in more detail, we have modeled the time development of the shadow diameter
as driven by the formation of the gravatom
numerically by solving Eqs.~\eqref{eq:spinevolutionsdiffeqs},~\eqref{eq:energyinstabilityflux},~\eqref{appeq:alphasky} 
and~\eqref{appeq:betasky} (see Appendix~\ref{app:BHshadowapp} for details).
For simplicity, we set the observer orientation to $\theta_o=\pi/2$.
As initial setup we choose scalar cloud seeds with
$M_{c,0}=10^{-9}M_{0}$ and $M_{c,0}=0.025 M_{0}$.
Exemplarily, we focus on black holes with their initial masses corresponding to the EHT's prime targets Sgr~A$^{\ast}$,
$M_{0}=4.2\times10^{6}M_{\odot}$,
and M87, $M_{0}=6.5\times10^{9}M_{\odot}$.
We fixed the gravitational coupling~\eqref{eq:DefGravCoupling}
to $\alpha = 0.05$.
That is, we probe for ultralight bosons with, respectively,
$\mu = 1.5\times10^{-18}$eV and $\mu=10^{-21}$eV.
We present the resulting evolution of the shadow's diameter $d_{\rm sh}$ for different initial spins in 
Figs.~\ref{fig:AngulardiameterevolutionSgrA}
and~\ref{fig:AngulardiameterevolutionM87}.

We observe that the angular diameter of the shadow increases during the superradiant evolution,
with the change depending on the initial black hole spin,
until it reaches a peak.
Afterwards it decreases and settles to a new (constant) value at the end of the superradiant evolution.
The peak is determined by
\begin{align}
\label{eq:dshEvolPean}
0 = \frac{\dif d_{\rm sh}}{\dif t} = &
        \frac{\dif M}{\dif t} \frac{\dif d_{\rm sh}}{\dif M} + \frac{\dif J}{\dif t} \frac{\dif d_{\rm sh}}{\dif J}
\nonumber \\ = &
        - 2 \Gamma_{nlm} M_{c} \left[ \frac{\dif d_{\rm sh}}{\dif M} + \frac{m}{\mu} \frac{\dif d_{\rm sh}}{\dif J} 
\right]
\,,
\end{align}
where we used Eqs~\eqref{eq:spinevolutionsdiffeqs} and~\eqref{eq:energyinstabilityflux}.
Let us now inspect the different terms more carefully. 
In phase I, i.e. during the superradiant evolution, the imaginary part of the frequency $\Gamma_{nlm}>0$, so for the shadow to grow the term in the brackets has to be negative.
This implies the condition
\begin{align*}
\frac{\mu}{m} \frac{\dif d_{\rm sh}}{\dif M} < & - \frac{\dif d_{\rm sh}}{\dif J}
\,,
\end{align*}
where $ \frac{\dif d_{\rm sh}}{\dif J} < 0$.
That is, the evolution of the shadow diameter is dominated by the spin-down of the central black hole 
rather than its change in mass.
At the peak itself there are two possibilities:
either $\Gamma_{nml}=0$ which corresponds to the end of the superradiant evolution 
or the term in the bracket of Eq.~\eqref{eq:dshEvolPean} vanishes, i.e.
\begin{align*}
\frac{\mu}{m} \frac{\dif d_{\rm sh}}{\dif M} = & - \frac{\dif d_{\rm sh}}{\dif J}
\,.
\end{align*}
Comparing the evolution of the shadow in Figs.~\ref{fig:AngulardiameterevolutionSgrA} and~\ref{fig:AngulardiameterevolutionM87}
to that of the gravatom in Fig.~\ref{fig:Spinevolutionseeds},
we see that the turning point corresponds to the latter condition.

Now the evolution of the shadow's angular diameter is dominated by its dependency on the mass, 
\begin{align*}
\frac{\mu}{m} \frac{\dif d_{\rm sh}}{\dif M} > & - \frac{\dif d_{\rm sh}}{\dif J}
\,.
\end{align*}
Since we are still in the superradiant regime~\eqref{eq:sprcond}, where $\Gamma_{nlm}>0$, relation~\eqref{eq:dshEvolPean}
indicates that the shadow diameter should decrease. This is indeed what we observe in Figs.~\ref{fig:AngulardiameterevolutionSgrA}
and~\ref{fig:AngulardiameterevolutionM87}.

This brief postpeak phase is succeeded by phase II, i.e., the new, quasistationary gravatom after the superradiant evolution. 
Here, the cloud is slowly decaying $\Gamma_{nlm}\lesssim 0$, whereas the black hole parameters no longer change and
its shadow remains (almost) constant as can be seen in Figs.~\ref{fig:AngulardiameterevolutionSgrA} and~\ref{fig:AngulardiameterevolutionM87}.

We further note that the significant exponential growth of the scalar cloud and superradiant reduction in the black hole mass and spin, 
phase I in Fig.~\ref{fig:Spinevolutionseeds},
takes place about $10^{7}\cdots10^{11}$ years after the beginning of the (superradiant) evolution. From here on we will refer to this as 
$t_{\rm I}$.
The precise value depends on the details of the superradiant evolution.
Note that here we neglected the accretion of ordinary matter. 
This process would move the black hole into the superradiant regime through the transfer of mass and 
angular momentum onto it, the process appears to not further impact the superradiant evolution itself on the relevant timescales~\cite{Brito:2014wla}.
We leave a more detailed study of this effect for future work.

\section{Analytic approximation}\label{sec:AnalyticApproximation}
As we have seen in the previous sections, the superradiant evolution as well as the angular diameter of the black hole shadow
are in general determined numerically. Although the tools fall into the realm of ``soft numerics,''
an analytic description would greatly enhance our understanding of the evolution
and enable us to cover a wider range of parameters at once.
Therefore, we have derived a set of fitting formulas for both 
the superradiant evolution and  the dependence of the black shadow on the black hole spin.
Details of the derivation can be found in Appendixes~\ref{app:BHspinevolutionapp} and~\ref{app:shadowkerrbh}.

\subsection{Modeling the superradiant evolution}\label{ssec:modelizingsuperradiance}
\begin{figure*}[htpb!]
\begin{center}
\subfloat[$M_{c,0}=10^{-9}M_{0}$]{\includegraphics[width=0.5\textwidth,clip]{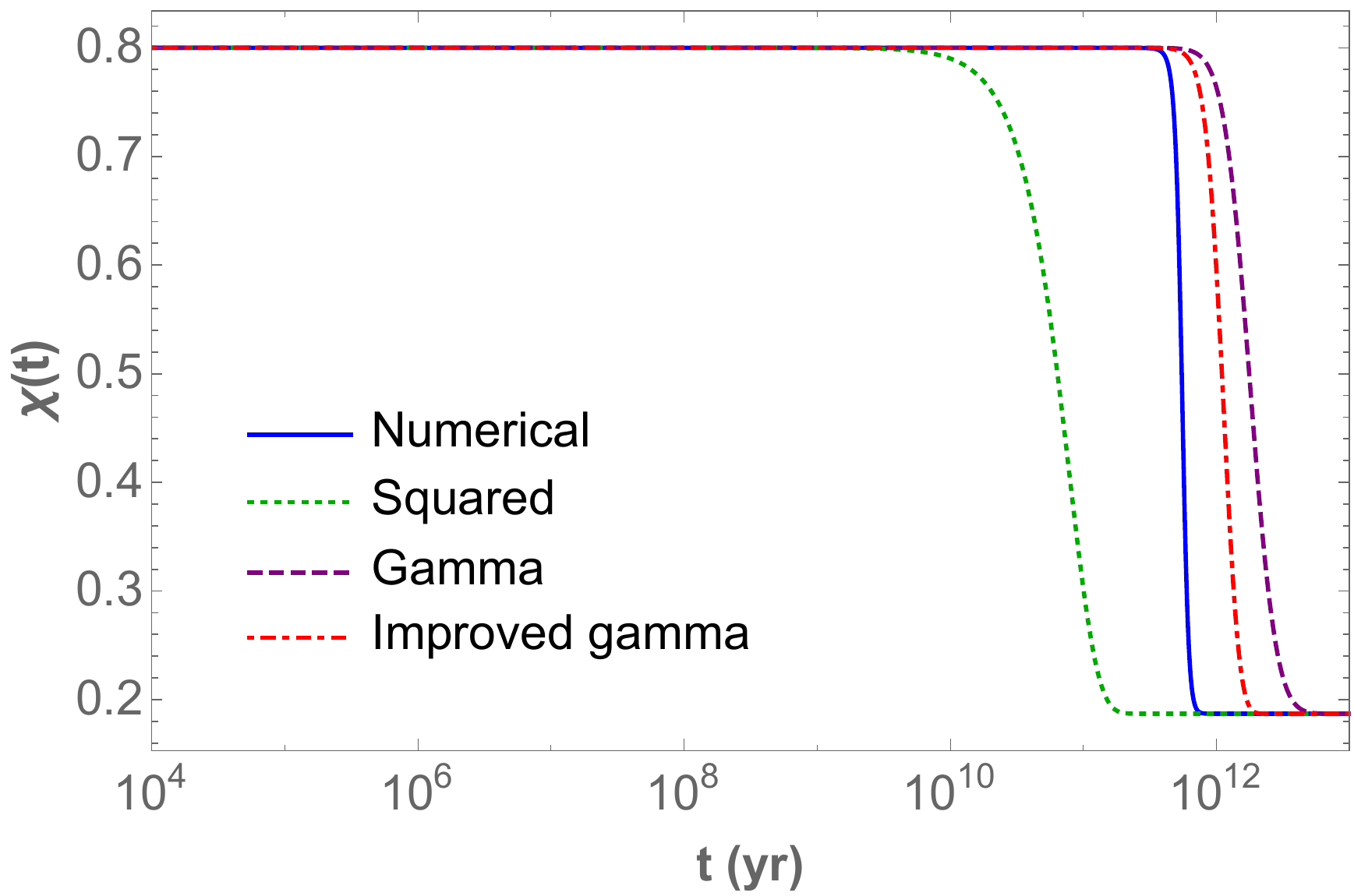}\label{fig:SpinevolutionM87small seed}}
\subfloat[$M_{c,0}=0.025M_{0}$]{\includegraphics[width=0.5\textwidth,clip]{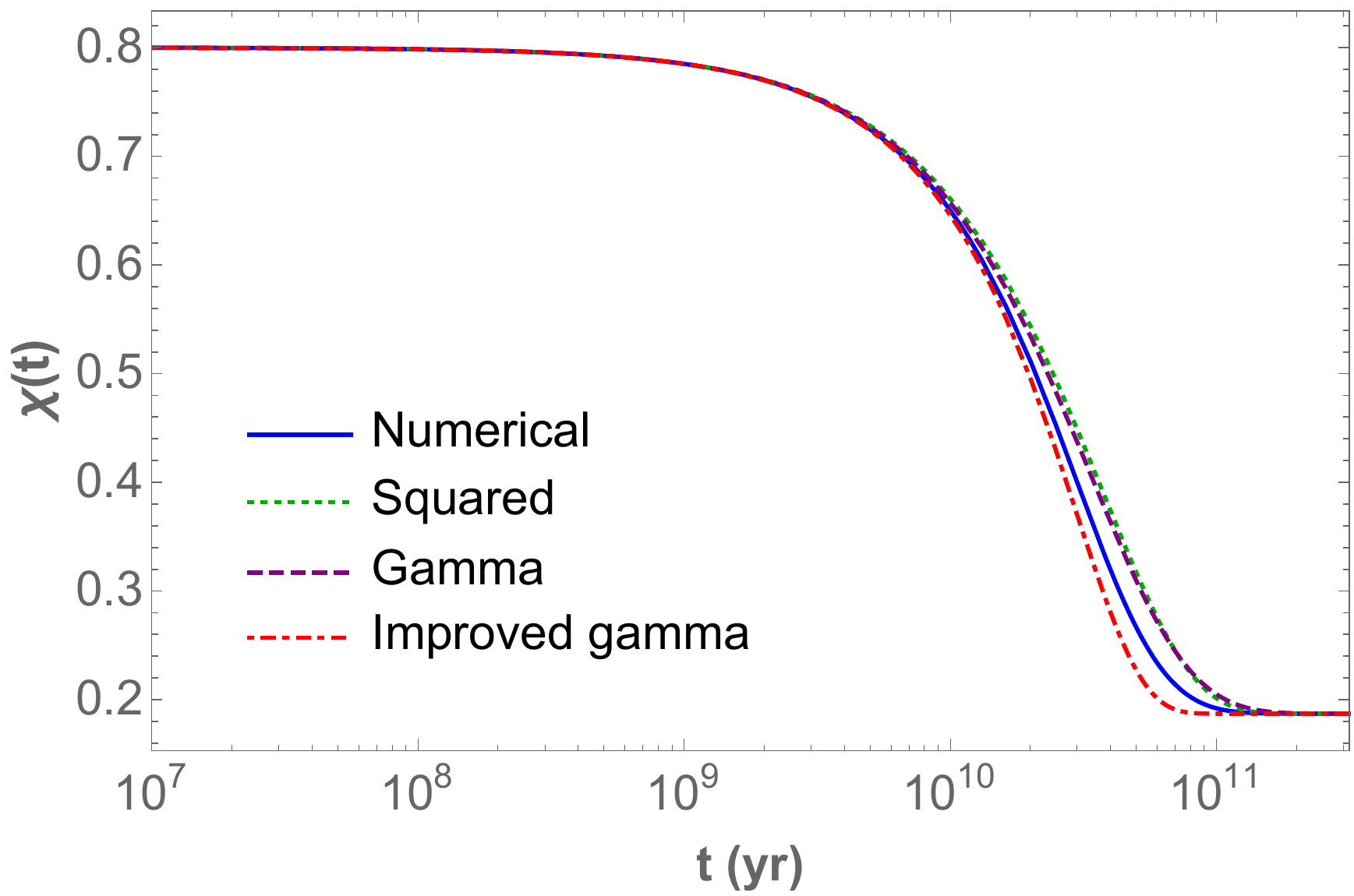}\label{fig:SpinevolutionM87largeseed}}
\caption{\label{fig:SpinevolutionM87}
        Evolution of the black hole spin $\chi$ for a M87-type black hole with initial mass $\MZ=6.5\times10^9M_{\odot}$ and initial spin $\chiZ=0.8$.
        The gravitational coupling is $\alpha=0.05$ so that the scalar condensate is composed of particles with 
        $\mu=1\times10^{-21}$eV. The initial cloud mass is $\McZ=10^{-9}\MZ$ (left) or $\McZ=0.025\MZ$ (right).
        We compare the performance of the different approximation methods to
        the adiabatic, i.e. numerical, evolution (solid blue line).
        For large initial seeds all fits reproduce the superradiant evolution reasonably well (right panel), 
        whereas approximating the evolution of small initial seeds is more challenging (left panel).
}
\end{center}
\end{figure*}

Here we derive approximate formulas that allow us to model the superradiant evolution
discussed in Sec.~\ref{sec:shadowsuperradiance}
{\textit{analytically}}.
The first ansatz, labeled {\textit{squared fit}}, attempts to emulate the time dependence of the black hole parameters
directly. Instead, the {\textit{gamma fit}} and {\textit{improved gamma fit}} promote the growth or decay rate $\Gamma$
to a time dependent quantity. 
Since the rate depends on the evolving black hole's mass and spin
the latter captures their time dependence.
The details of the derivation are given in App.~\ref{app:BHspinevolutionapp}.
In Fig.~\ref{fig:SpinevolutionM87} we present a comparison between the numerically computed adiabatic evolution and the different approximation schemes.
In particular, we present the time evolution of the dimensionless black hole spin for small and large initial scalar clouds. 
The fits appear to perform better for large initial clouds, where the time gradients are less steep. Overall,
the improved gamma fit performs best.
In Appendix~\ref{app:BHspinevolutionapp} we include an additional approximation that performs well for small seeds,
but needs to be fine-tuned for the specific initial conditions.

\subsubsection{Squared fit}
The squared fit consists of an exponential ansatz for the black hole's and the scalar cloud's parameters
and has the form
\begin{align}
\label{eq:AnsatzSquaredFit}
p(t) = & p_{\rm II} + \left(p_{0} - p_{\rm II} \right)\exp\left[-\gamma t - \beta t^2 \right]
\,,
\end{align}
where $p$ is a placeholder for the time dependent quantities $(M, J, M_{c}, J_{c})$,
and the subscripts $_{0}$ and $_{\rm II}$ refer to their values at the beginning (phase 0) and end (phase II) of the superradiance instability (see Sec.~\ref{sec:Evolution}), respectively.
The exponents are determined by
\begin{align}
\label{eq:SquaredFitgammabeta}
\gamma = & 2\Gamma_{0} \frac{M_{c,0}}{M_{0}-\MII}
\,,\quad
\beta  = \frac{\ln\,2}{t^{2}_{\ast}} - \frac{\gamma}{t_{\ast}}
\,,
\end{align}
where $\Gamma_{0}\equiv\Gamma(t=0)$ is the initial growth rate
given in Eq.~\eqref{eq:SFFreqIm} and 
\begin{align}
t_{\ast} = & \frac{2}{\gamma} \frac{M_{c,0}}{M_{c,0}+M_{c,\rm II}}
\,,
\end{align}
refers to the time when the system's variables have reached their mean value
$p_{\ast} \equiv p(t_{\ast}) = \frac{p_{0}+p_{\rm II}}{2}$.

As we see in Fig.~\ref{fig:SpinevolutionM87}, 
this approximation seems to perform well if the scalar cloud seed is 
of a few percent of the black hole mass. 
Small seeds, however, are not modeled well, and one would have to include higher powers in $t$ to capture the steeper
gradients during their evolution. Technically, this approach would render the system of equations underdetermined for a 
finite number of additional powers.

\subsubsection{Gamma fit}\label{Gammafit}
To circumvent these shortcomings of the squared fit, the following approximations follow a different avenue: 
we consider the exponential of an
exponential (instead of a power-law) function, and we promote the growth rate $\Gamma$ to a time dependent variable.
In particular, we take the ansatz
\begin{align}
\label{eq:AnsatzGammaFit}
\Gamma(t) = & \Gamma_0 \exp\left[-\gamma{t} \right]
\,,
\end{align}
where the initial growth rate $\Gamma_{0}$ is given in Eq.~\eqref{eq:SFFreqIm}.
Substituting this ansatz into the evolution equations~\eqref{eq:spinevolutionsdiffeqs} and~\eqref{eq:energyinstabilityflux}
we obtain
\begin{align}
\label{eq:GammaFitEvolution}
M_{c} (t) = & M_{c,0}\exp\left[\frac{2\Gamma_{0}}{\gamma}\left(1-e^{-\gamma{t}}\right)\right]
\,,\\
M(t) = & M_{0} - M_{c,0} \left\{\exp\left[\frac{2\Gamma_{0}}{\gamma} \left(1-e^{-\gamma\,t}\right) \right] - 1 \right\}
\,,\nonumber\\
J(t) = & J_{0} - M_{c,0} \frac{m}{\mu} \left\{\exp\left[\frac{2\Gamma_{0}}{\gamma} \left(1-e^{-\gamma\,t}\right) \right] - 1 \right\}
\,,\nonumber
\end{align}
and the exponent reads
\begin{align}
\label{eq:gammaGammafit}
\gamma = & \frac{2\Gamma_0}{\ln\left({\frac{M_{c,\rm II}}{M_{c,0}}}\right)}
\,.
\end{align}
As illustrated in Fig.~\ref{fig:SpinevolutionM87}, the gamma fit is comparable to the squared fit for large scalar cloud seeds,
but seems to perform significantly better for small seeds.

\subsubsection{Improved gamma fit}
We now propose a more sophisticated ansatz for the superradiance rate given by
\begin{align}
\label{eq:improgammat}
\Gamma(t) = & \Gamma_0 \exp\left[1-e^{\gamma{t}}\right]
\,,
\end{align}
where $\Gamma_{0}$ is the initial growth rate; see Eq.~\eqref{eq:SFFreqIm}.
The evolution of the scalar cloud is described by
\begin{align}
\label{eq:ansatzMcimp}
M_c(t) = & M_{c,0} \exp\left[\frac{2\Gamma_0e}{\gamma}\left(E[-e^{\gamma{t}}]-E[-1]\right)\right]
\end{align}
where the exponent is
\begin{align}
\label{eq:improgamma}
\gamma = & -\frac{2\Gamma_0E[-1]e}{\ln\left({\frac{M_{c,\rm II}}{M_{c,0}}}\right)}
\,,
\end{align}
and $E[x]$ the exponential integral is defined as
\begin{align*}
E[x] = & -\int_{-x}^{\infty}\frac{e^{-t}}{t}dt
\,.
\end{align*}
In particular, $E[-1]=-0.219384$. 
The evolution of the black hole's mass and spin is determined by
\begin{subequations}
\label{eq:ImprovedGammaEvolution}
\begin{align}
\label{eq:ansatzMimp}
M(t) = & M_0 - M_{c,0}\left\{\exp\left[\frac{2\Gamma_0e}{\gamma}\left(E[-e^{\gamma{t}}]-E[-1]\right)\right]-1\right\}
\,,\\
\label{eq:ansatzJimp}
J(t) = & J_0 - \frac{m}{\mu}M_{c,0}\left\{\exp\left[\frac{2\Gamma_0e}{\gamma}\left(E[-e^{\gamma{t}}]-E[-1]\right)\right]-1\right\}
\,.
\end{align}
\end{subequations}
The improved gamma fit performs better than the squared or gamma fit; see Fig.~\ref{fig:SpinevolutionM87}.
However, it still involves solving the exponential integral numerically.

\subsubsection{Regime of validity}
We compare the different approximation schemes to the numerical computation 
in Fig.~\ref{fig:SpinevolutionM87}
exemplarily for the evolution of the black hole spin.
This figure focuses on a specific set of initial parameters, namely $\chiZ=0.8$ and $\alpha=0.05$.
Here we investigate the range of validity of fits in more detail. 
We focus on the gamma and improved gamma fits that reproduce the numerical evolution best.

We vary the initial black hole spins $\chiZ = 0.5,\ldots,0.99$
that are representative values well within the superradiant regime~\eqref{eq:sprcondSpin},
and we vary the gravitational coupling between $\alpha=0.01,\ldots 0.1$.
The maximal value denotes the breakdown of the small-coupling approximation.
At that point the numerical evolution itself becomes invalid.
We concentrate on small and large scalar cloud seeds as extreme cases in the $\McZ$ range.

We find that the error is generally smaller for smaller values of the initial black hole spin. The smaller the gravitational coupling, the smaller is the relative error in the black hole mass but the larger the error in the black hole spin.
In the case of the black hole mass $M(t)$, the analytic approximation agrees with the numerical
data within $\lesssim1\%$ ($\lesssim7\%$) for large (small) scalar cloud seeds and the entire 
spin-coupling parameter range.
The evolution of the spin, however, is more sensitive to the chosen parameters. 
While the approximation is not valid for small scalar cloud seeds, 
it describes the spin evolution within better than $\sim 20\%$ ($\alpha=0.05$, $\chiZ=0.99$) 
and better than $\sim 15\%$ ($\alpha=0.1$, $\chiZ=0.99$). 
The deviation reaches $\lesssim10\%$ for spins $\chiZ\lesssim0.8$ and couplings $\alpha\gtrsim0.05$.

\subsection{Modeling the shadow}\label{ssec:blackholeshadow}
\begin{figure}[h!]
\begin{center}
\includegraphics[width=0.475\textwidth,clip]{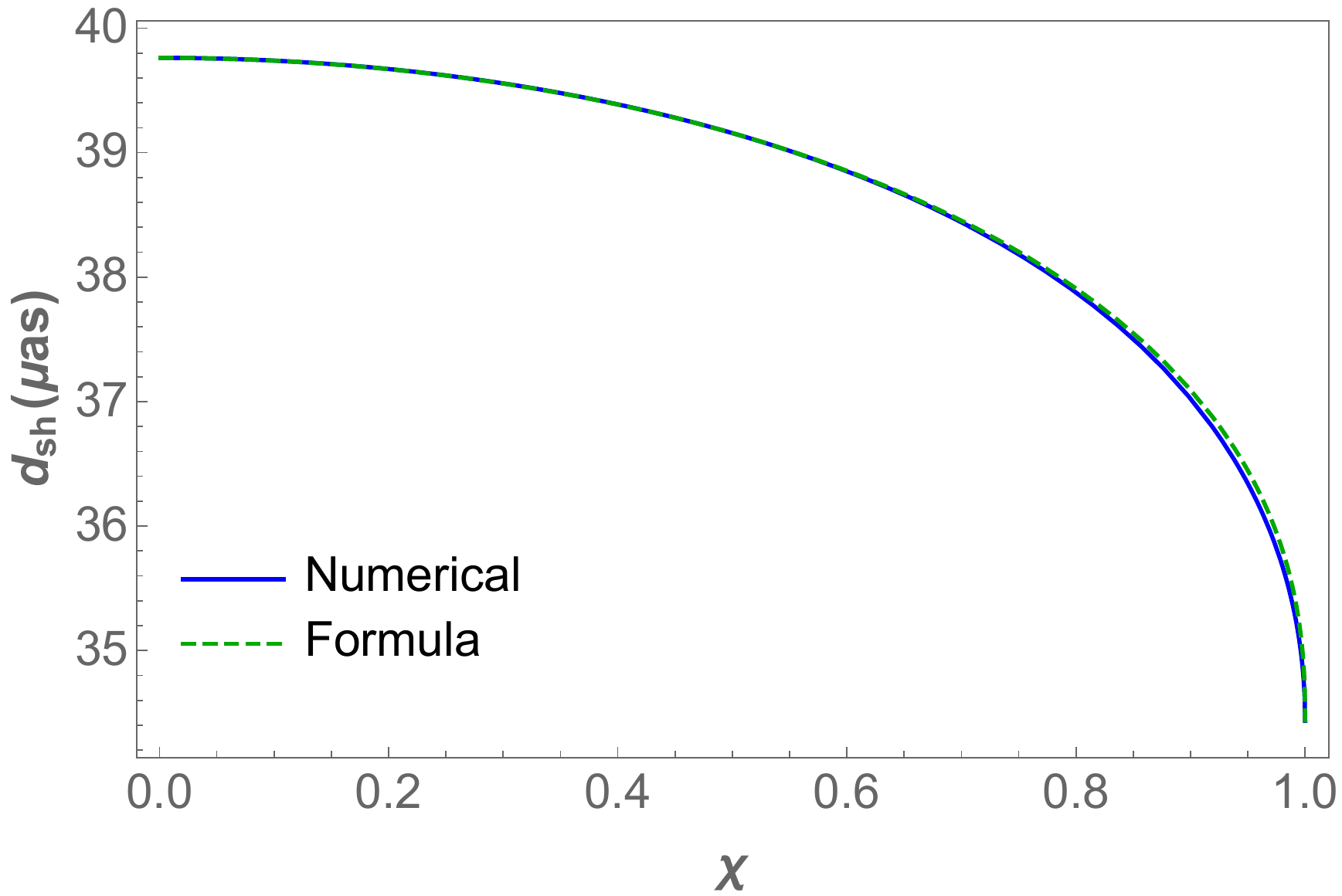}
\caption{\label{fig:AngulardiameterBHshadowdsh}
        Shadow's angular diameter $d_{\rm sh}$ as a function of the dimensionless black hole spin $\chi$
        for a black hole of mass $M_{\rm M87}=6.5\times10^{9}M_{\odot}$ at a distance $r_{o}=16.8$Mpc.
        We compare the analytic approximation~\eqref{eq:shadowdiameteranalytical} (green dashed line)
        to its numerical evaluation (blue solid line) and find agreement within $\lesssim0.5\%$.
}
\end{center}
\end{figure}

\begin{figure}[h!]
\begin{center}
\includegraphics[width=0.475\textwidth,clip]{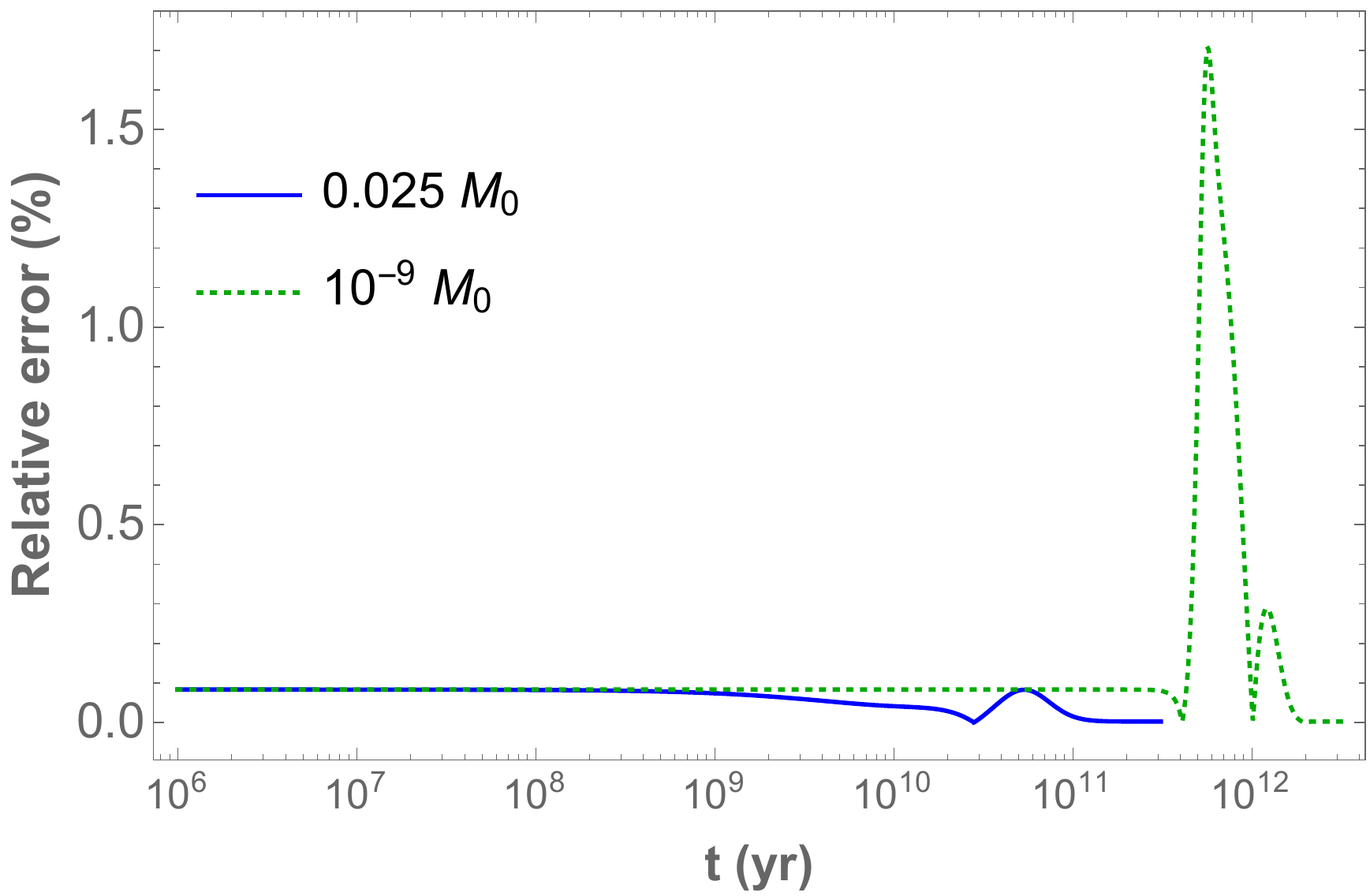}
\caption{\label{fig:BHShadowEvolNumericalVSImprovedGammaFit}
Relative error of the shadow diameter $d_{\rm sh}$ as a function of time 
obtained with the approximation~\eqref{eq:shadowdiameteranalytical}
as compared to the numerical evolution.
We set the initial black hole mass $\MZ=6.5\times10^{9}M_{\odot}$ and spin $\chiZ=0.8$, and the coupling $\alpha=0.05$.
The initial cloud mass is $\McZ=10^{-9}\MZ$ (green dotted curve) or $\McZ=0.025\MZ$ (blue solid curve). 
The end of the solid curve corresponds to the end of the superradiant evolution.
}
\end{center}
\end{figure}

Relating the black hole shadow to its spin is, in general, a nontrivial task that needs to be solved numerically.
Here we derive an analytic formula, fitted to our numerical results presented in Sec.~\ref{sec:shadowsuperradiance}.
Specifically, we take a power-law ansatz of the form
\begin{align}
\label{eq:ansatzdshVsChi}
d_{\rm sh} = & A + B \left( 1 - \chi^2 \right)^{\delta}
\,,
\end{align}
where the choice $\chi^2$ enforces the symmetry under $\chi\rightarrow -\chi$.
We determine the parameters $(A,B,\delta)$ by evaluating the full expression for three different values of the 
dimensionless spin under the simplifying assumption that the observer is located in the equatorial plane; 
see Appendix~\ref{app:shadowkerrbh}.
The coefficients are
\begin{subequations}
\label{eq:coefficientsdshVsChi}
\begin{align}
A = & \frac{9M}{r_{o}}
\,,\quad
B =   \frac{3 \left(2\sqrt{3}-3\right) M}{r_{o}}
\,,\\
\delta = & \frac{\ln(\frac{2\sqrt{3}-3}{2\sqrt{3}S-3})}{\ln(4/3)}\sim 0.4
\,,
\end{align}
\end{subequations}
where $M$ is the black hole mass, $r_{o}$ is its distance to the observer and we introduced 
$S = \sin \frac{\pi}{9} + \sin \frac{2\pi}{9}$.
Then, the shadow diameter can be approximated as 
\begin{align}
\label{eq:shadowdiameteranalytical}
d_{\rm sh} = & \frac{3M}{r_o}\left[3+(2\sqrt{3}-3)\left(1-\chi^2\right)^\delta\right]
\,.
\end{align}
In Fig.~\ref{fig:AngulardiameterBHshadowdsh}
we compare this analytic formula with our numerical data. We find excellent agreement within $\lesssim0.5\%$
for high spins and better for small spins.
We can now directly relate the measured shadow diameter to the black hole spin
(assuming we have determined the black hole mass and distance to the observer independently)
by inverting Eq.~\eqref{eq:shadowdiameteranalytical}.
We find
\begin{align}
\label{eq:diametertospin}
\chi = & \pm\sqrt{1 - \left(\frac{\frac{r_o d_{sh}}{3M}-3}{2\sqrt{3}-3}\right)^{1/\delta}}
\,.
\end{align}
Furthermore, we approximate the development of the shadow diameter $d_{\rm sh}$ due to the superradiant evolution by promoting $M \rightarrow M(t)$ and $\chi \rightarrow \chi(t)$ in Eq.~\eqref{eq:shadowdiameteranalytical}.
We model them with the improved gamma fit, Eqs.~\eqref{eq:ImprovedGammaEvolution},
since they best approximate the gravitational atom.
In Fig.~\ref{fig:BHShadowEvolNumericalVSImprovedGammaFit}
we compare the numerical and analytic data for the evolution of the shadow diameter of a black hole 
with an initial mass $\MZ\sim M_{\rm M87}=6.5\times10^{9}M_{\odot}$ and spin $\chiZ=0.8$
for small and large scalar cloud seeds.
The shadow diameter obtained with the analytic approximation agrees within
$\lesssim2\%$ ($\lesssim0.5\%$) for small (large) seeds with the numerical computation;
cf.~Fig.~\ref{fig:BHShadowEvolNumericalVSImprovedGammaFit}.
The magnitude of the uncertainty, especially when compared to deviation of the spin evolution shown in 
Fig.~\ref{fig:SpinevolutionM87}
can be understood by studying the propagation of errors. Applying it to Eq.~\eqref{eq:shadowdiameteranalytical}
and inserting the uncertainty in the black hole mass and spin quoted in the previous section,
we find a relative error of a few percent.
That is, it is consistent with the direct comparison shown in Fig.~\ref{fig:BHShadowEvolNumericalVSImprovedGammaFit}.

\section{Imaging the superradiant evolution}
The shadows of black holes with synchronized scalar hair -- that is, the final state of the superradiant evolution of complex ultralight bosons -- have received broad attention in the literature. In particular,
tracing out the light rays around such solutions revealed an intricate structure of their shadows~\cite{Cunha:2015yba,Cunha:2016bpi,Roy:2019esk}
although the recent observations of M87 by the EHT place only weak constraints on these
spacetimes~\cite{Cunha:2019ikd}.

Here we focus on a different question: can one detect 
the formation of these hairy black holes,
i.e., can one ``record'' the superradiant evolution itself within a few decades of observations?
To address this question
we focus on phase~I of the evolution, (cf. Fig.~\ref{fig:Spinevolutionseeds}) during which the black hole parameters and, consequently, the shadow diameter undergo the largest changes.
In this section we investigate three items in particular:
\begin{enumerate*}[label={(\roman*)}]
\item How long would it take to reach phase~I?
\item Assume we can measure a change in the shadow; how can we infer the boson's mass?
\item How does the shadow change?
\item How large would this change be over relevant observation timescales?
\end{enumerate*}
Unless stated otherwise, we assume an observation time of about $30$ years.

\subsection{Time to reach phase~I}\label{ssec:ReachingPhaseI}
The largest changes in the shadow occur during phase~I (see Fig.~\ref{fig:Spinevolutionseeds}),
so it provides the best-case scenario to detect (or constrain) the superradiant evolution. 
To estimate the time $\tI$ it would take to reach this stage after the onset of the superradiant instability, 
we start from the definition of $\tI$ given in Sec.~\ref{sec:Evolution} and use the gamma fit approximation 
presented in Sec.~\ref{Gammafit}. 
Starting from Eq.~\eqref{eq:GammaFitEvolution} we find 
\begin{align}
\tI = -\frac{\ln\left(\frac{M_{c,f}}{M_{c,0}}\right)}{2\Gamma_{0}}\ln\left(1-\frac{\ln\left(\mathcal{Z}\frac{M_0}{M_{c,0}}+1\right)}{\ln\left(\frac{M_{c,f}}{M_{c,0}}\right)}\right)
\,.
\end{align}
Here, $\McII$ is determined by $\MII=M_0+M_{c,0}-M_{c,\rm II}$ and $\MII$ by Eq.~\eqref{eq:finalmassBH}.
We find results consistent with the numerics if we choose ${\cal Z}=10^{-4}$ (${\cal Z}=10^{-5}$) for large (small) scalar seeds.
For example, for M87 and SgrA$^{\ast}$ (and exemplarily setting $\chi_0=0.8$), we get
$\tI^{\rm M87}{}_{0.025M_0}=1\times10^8$ yr, 
$\tI^{\rm M87}{}_{10^{-9}M_0}=3.3\times10^{11}$ yr 
and $\tI^{\rm SgrA*}{}_{0.025M_0}=8.7\times10^4$ yr, 
$\tI^{\rm SgrA*}{}_{10^{-9}M_0}=2.9\times10^8$ yr. 
The time to reach phase~I depends on the mass of the seed and the black hole. Generically, small seeds lead to larger $\tI$.
For larger seeds, of the order of 2.5\% of the initial black hole mass, one finds timescales varying between a few ten thousand years and 100 million years for supermassive black holes such as SgrA* and M87 respectively. For stellar size black holes, this can be much faster, of the order of magnitude of a year or even months for large seeds. 
In Table~\ref{tab:tablesummaryBHexample} we present concrete values for the coupling $\alpha=0.05$.
Larger values of the gravitational coupling constant seem to reduce the timescales substantially, 
but the small-coupling approximation is less reliable.

\begin{figure}[h!]
\begin{center}
\includegraphics[width=0.45\textwidth,clip]{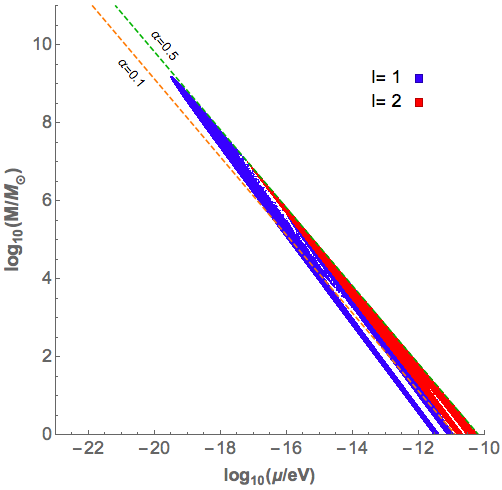}
\caption{\label{fig:Observationparameterspacetobsgammaimpr}
Configurations in the black hole mass--boson mass phase space that could reach phase~I within $t_{I}\lesssim10$ yr.
The collection of curves corresponds to gravitational coupling in the range $0<\alpha<0.5$ and initial black hole spin $0.5\leq\chi_{0}\leq0.99$ for both small and large initial scalar seeds.
We considered scalars composed of an $l=m=1$ mode (blue curves) and $l=m=2$ mode (red curves).
We denote $\alpha=0.5$ (green dashed line) and the (approximate) small coupling regime $\alpha\lesssim0.1$ (orange dotted line).
}
\end{center}
\end{figure}

To reconnect to observations of the black hole shadow, let us assume that the EHT (or a follow-up mission thereof) is operational for the next couple of decades.
Are there any gravitational atoms that would form within that time window, and what would their parameters be?
This is illustrated in Fig.~\ref{fig:Observationparameterspacetobsgammaimpr},
where we show all configurations in the black hole mass $M$--boson mass $\mu$ phase space that would reach phase~I
within $\tI\lesssim10$ yr after the onset of the instability.
The plot depicts a collection of curves that corresponds to broad ranges in parameter space.
Specifically, we varied the gravitational coupling  $0<\alpha<0.5$ and the initial black hole spin $0.5\leq\chi_{0}\leq0.99$, included small- and large-seed scalar field initial data and considered scalars with an initial $l=m=1$ mode (blue curves) or $l=m=2$ modes (red curves) 

As can be seen in Fig.~\ref{fig:Observationparameterspacetobsgammaimpr}, in principle one might be able to probe for
the formation of bosonic condensates of mass parameter $10^{-20}\lesssim \mu/{\rm eV} \lesssim 10^{-11}$ for black holes in the range $1\lesssim M/M_{\odot} \lesssim 10^{9}$. More massive black holes would require a longer time $\tI$ to reach phase~I.

\subsection{Measuring the boson mass}\label{ssec:MeasureMu}
We now derive a relation that allows us to estimate 
the gravitational coupling $\alpha$
from the measured change in the shadow diameter $d_{\rm sh}$.
To simplify the derivation, we fix the observer's orientation and distance $r_{o}$ to the source.
Since the superradiant evolution affects the black hole mass $M$ and spin $J$, we can write the shadow's change in time as
[cf. Eq.~\eqref{eq:dshEvolPean}]
\begin{align}
\label{eq:dshdtChainRule}
\frac{\dif d_{\rm sh}}{\dif t} = & \frac{\dif M}{\dif t} \left( \frac{\p d_{\rm sh}}{\p M} + \frac{m}{\mu}  \frac{\p d_{\rm sh}}{\p J} \right)
\,,
\end{align}
where we used Eq.~\eqref{eq:spinevolutionsdiffeqs}.
We use the shadow fitting formula, Eq.~\eqref{eq:shadowdiameteranalytical},
to derive 
\begin{subequations}
\label{eq:dshdM_dshdJ}
\begin{align}
\frac{\p d_{\rm sh}}{\p M} = & \frac{d_{\rm sh}}{M} + \frac{12\,\delta\,\chi^2}{r_{o}}\left(2\sqrt{3}-3\right) \left(1-\chi^2 \right)^{\delta-1}
\,,\\
\frac{\p d_{\rm sh}}{\p J} = & - \frac{6\,\delta\,\chi}{M\,r_{o}} \left(2\sqrt{3}-3\right) \left(1-\chi^2 \right)^{\delta-1}
\,,
\end{align}
\end{subequations}
where 
$\chi=J/M^2$ and
the exponent is 
$\delta = \frac{\ln(\frac{2\sqrt{3}-3}{2\sqrt{3}S-3})}{\ln(4/3)}$
with
$S = \sin \frac{\pi}{9} + \sin \frac{2\pi}{9}$.
Let us assume that the black hole parameters, the variation of the black hole mass,
$\frac{\dif M}{\dif t}$, and the change $\frac{\dif d_{\rm sh}}{\dif t}$ of its shadow can be measured independently.
Then, the gravitational coupling is determined by
\begin{align}
\label{eq:MeasureMuCase1}
\frac{\alpha}{m} = \frac{M\mu}{m} = & \frac{ M \p d_{\rm sh} / \p J }{ \left( \dif d_{\rm sh}/\dif t\right) \left(\dif M / \dif t \right)^{-1} - \p d_{\rm sh} / \p M 
}
\,,
\end{align}
as follows from Eq.~\eqref{eq:dshdtChainRule}
and where the coefficients are given in Eqs.~\eqref{eq:dshdM_dshdJ}. This provides a way to measure the boson mass $\mu$ and by multiplying with $M$, the gravitational coupling $\alpha$.

\subsection{Observing evolving shadows}\label{ssec:ShadowObs}
To understand better the observational prospects, we here investigate the magnitude of the changes in the black hole shadow due to the superradiant evolution.
We focus on representative stellar-mass and supermassive black holes, in particular
\begin{enumerate*}[label={(\roman*)}]
\item Cygnus X-1, historically the first black hole candidate;
\item GW170729, one of the most massive gravitational events in LIGO-Virgo's second observation run O2~\cite{LIGOScientific:2018mvr};
\item Sgr A$^{\ast}$, the supermassive black hole at the center of the Milky Way;
and
\item M87, the supermassive black hole whose shadow has been observed with the EHT.
\end{enumerate*}
We summarize their properties
and resulting shadow parameters in Table~\ref{tab:tablesummaryBHexample}, where we determine for instance $\Delta{d_{\rm sh}} $, the change in the shadow from $t_{I}$, the start of phase I, to $t_{I}+30$yr.
We focus on each type, stellar-mass and supermassive black holes, in more detail. 

\begin{table}[htpb!]
\begin{center}
\caption{\label{tab:tablesummaryBHexample}
Summary of black hole parameters mass $M$, dimensionless spin $\chi$, distance $r_{o}$, orientation $\theta_{o}$ and their angular resolution $R$.
To determine the changes $\Delta d_{\rm sh}$ and the timescale $t_{I}$ to reach phase~I we fixed the gravitational coupling $\alpha=0.05$
which corresponds to values of the boson mass $\mu$ given in the table.
The superscripts $\mathcal{L}$ ($\mathcal{S}$) refer to large (small) scalar cloud seeds 
The orientation of GW170729 and Sgr~A$^{\ast}$ are unclear, so we exemplarily set it to $\pi/2$.
}
\begin{tabular}{c|c|c|c|c}
                                  & Cyg-X1                        & GW170729            & Sgr~A$^{\ast}$                  & M87 \\
\hline
$M (M_{\odot})$                   & $14.8$                        & $80$                & $4.2\times10^{6}$               & $6.5\times10^{9}$ \\
$\chi$                            & $0.95$~\cite{Gou:2011nq}      & $0.81$              & $0.65$~\cite{Dokuchaev:2013xda} & $0.9$~\cite{Tamburini:2019vrf} \\
$r_{o}$                           & $1.9$kpc                      & $2750$Mpc           & $8.2$kpc                        & $16.8$Mpc \\
$\theta_{o}$                      & $3\pi/20$                     & $\pi/2$             & $\pi/2$                         & $17\pi/180$ \\
$R\,(\log_{10}[\mu\textrm{as}])$  & $-4$                          & $-9$                & $+1$                            & $+1$\\
$\mu$ [eV]                        & $5\times10^{-13}$           & $8\times10^{-14}$   & $1.5\times10^{-18}$             & $10^{-21}$\\
\hline
$t^{\mathcal{L}}_{I}$ [yr]               & $0.2$                         & $1.2$               & $8\times10^{5}$                 & $8.7\times10^{7}$\\
$\Delta d^{\mathcal{L}}_{\rm sh}$ [$\mu$as] & $2\times10^{-5}$              & $1.5\times10^{-11}$ & $2\times10^{-6}$                & $3\times10^{-9}$ \\
\hline
$t^{\mathcal{S}}_{I}$ [yr]               & $591$                         & $4040$              & $2.8\times10^{8}$               & $2.8\times10^{11}$ \\
$\Delta d^{\mathcal{S}}_{\rm sh}$ [$\mu$as] & $1\times10^{-6}$              & $3\times10^{-13}$   & $4\times10^{-8}$                & $5\times10^{-11}$\\
\hline
\end{tabular}
\end{center}
\end{table}

\noindent{\textbf{Stellar-mass black holes:}}
The angular resolution necessary to resolve a black hole's shadow is roughly determined by the ratio $M/r_{o}$ between its mass and distance to the observer; see Eq.~\eqref{eq:shadowdiameteranalytical}.
For example, Cyg X-1, a black hole candidate of about $15M_{\odot}$ in our galactic neighborhood at a distance of about $1.9$kpc would require an angular resolution of $\sim 8\times10^{-4}~\mu$as.
This is out of reach for the EHT. Therefore, observing the shadow of stellar-mass black holes
-- let alone its evolution -- is not feasible.

\noindent{\textbf{Supermassive black holes:}}
Therefore, let us focus on supermassive black holes.
They have two advantages: 
their shadow diameter is sufficiently large to be observable by the EHT 
as was shown by the images of M87,
and our results in Sec.~\ref{ssec:ShadowEvolution}
indicate that the change of the shadow diameter during the entire superradiant evolution 
is of the order of a few $\mu$as so may be detectable with future very long baseline interferometry (VLBI) instruments.

\begin{figure*}
\begin{center}
\includegraphics[width=.28\textwidth]{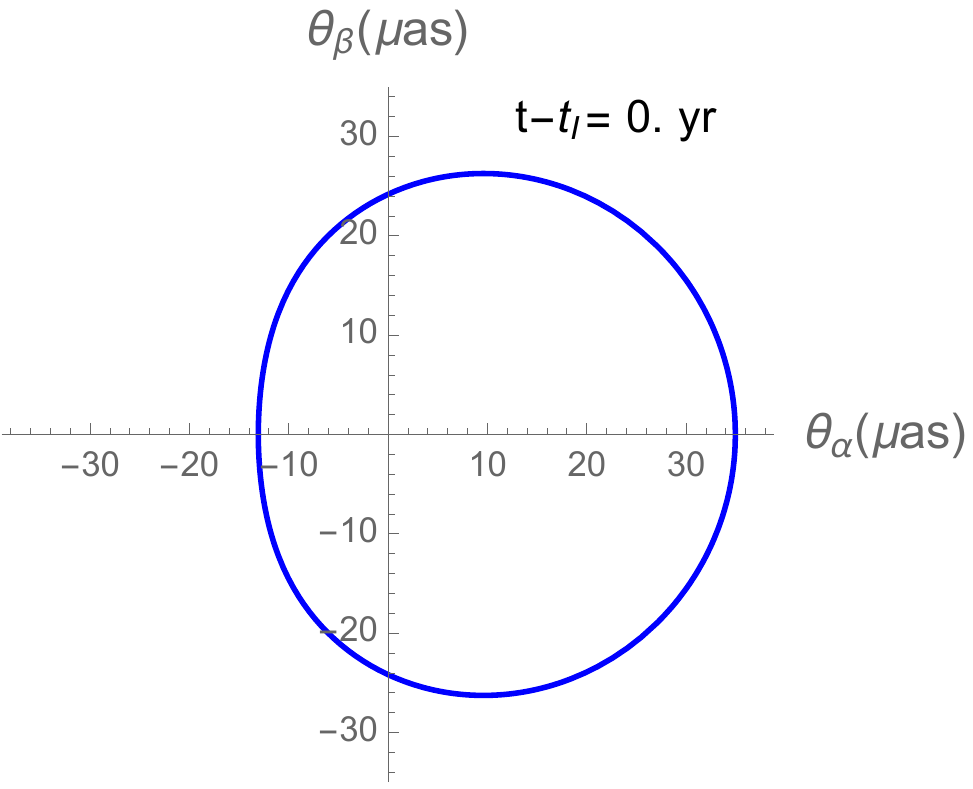}
\includegraphics[width=.28\textwidth]{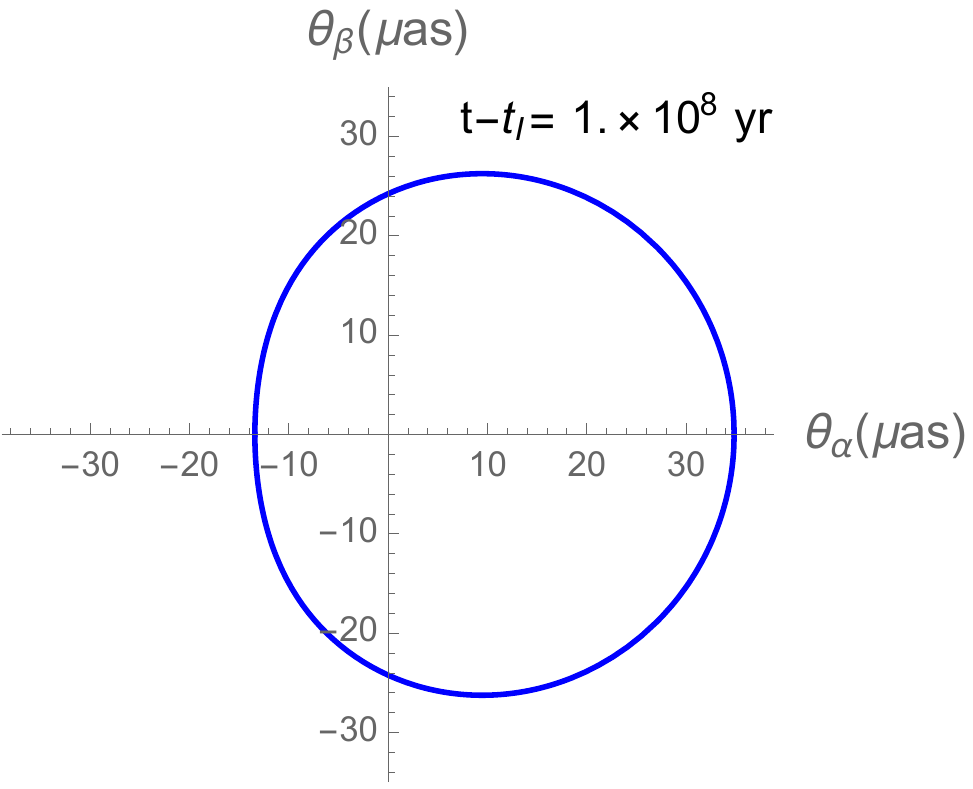}
\includegraphics[width=.28\textwidth]{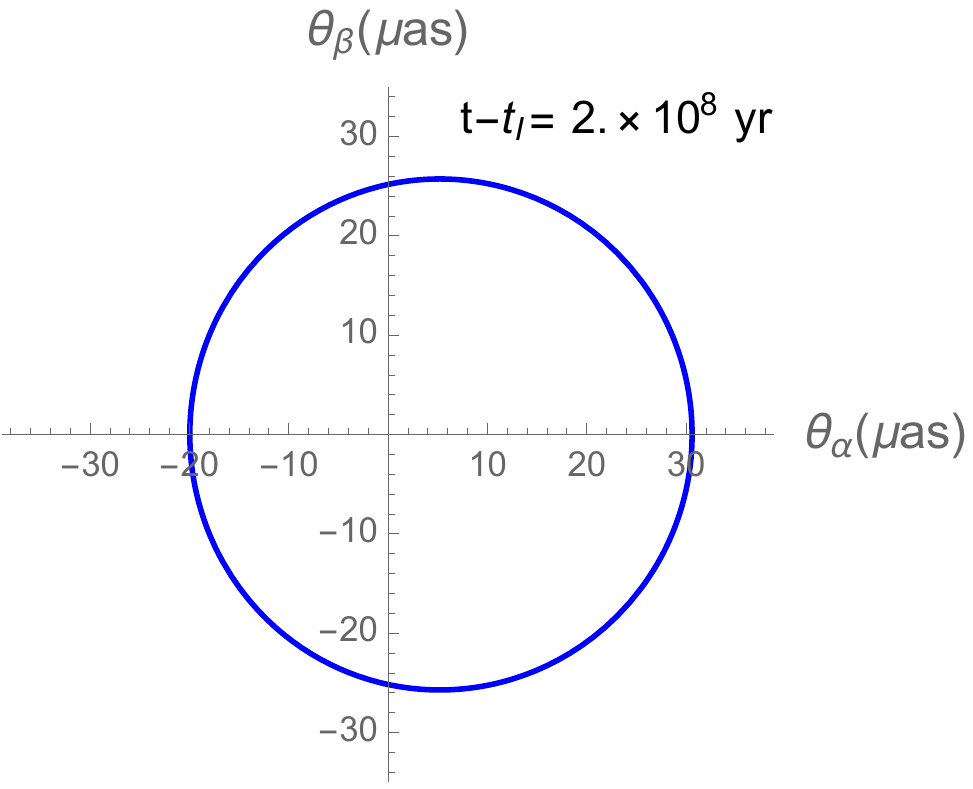}
\includegraphics[width=.28\textwidth]{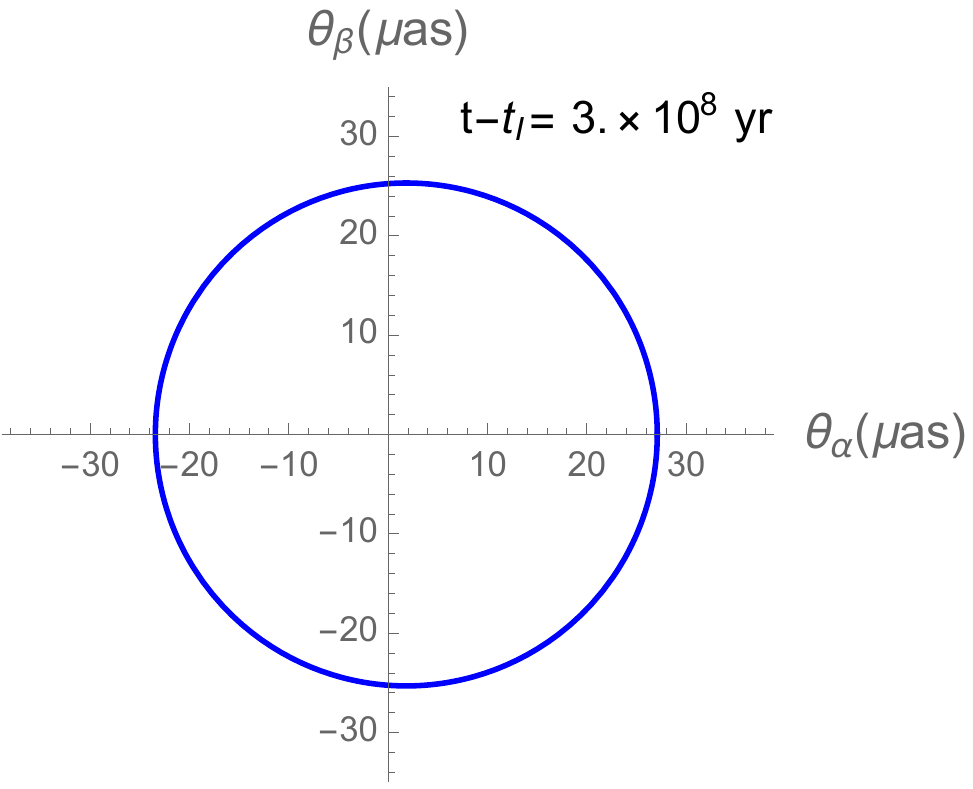}
\includegraphics[width=.28\textwidth]{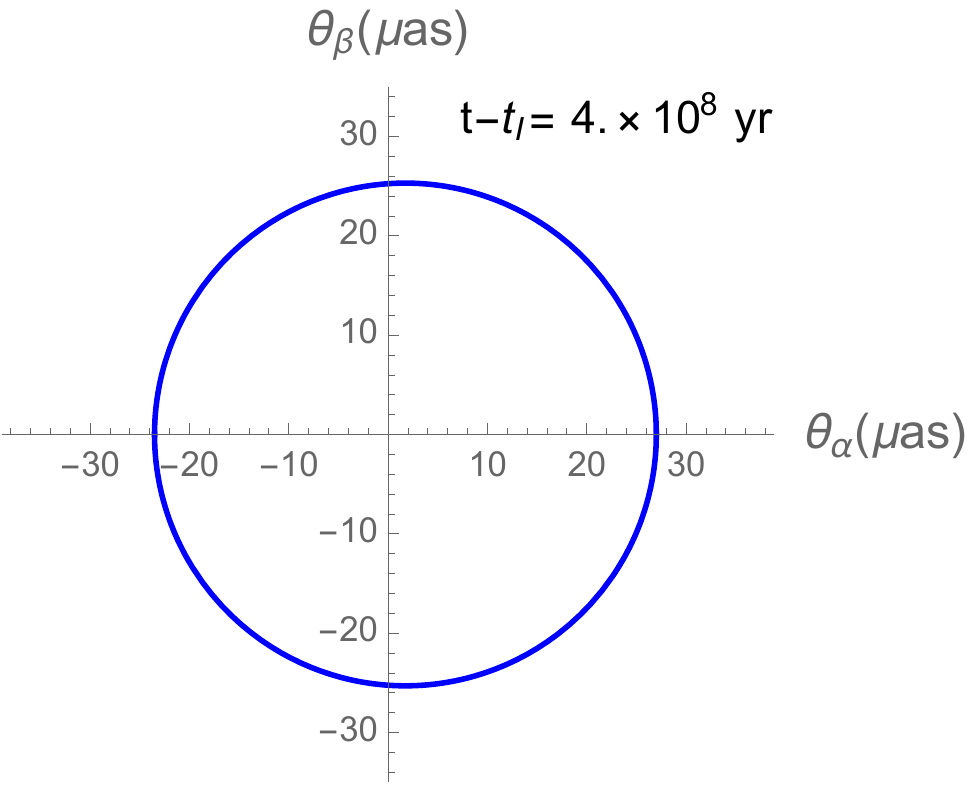}
\includegraphics[width=.28\textwidth]{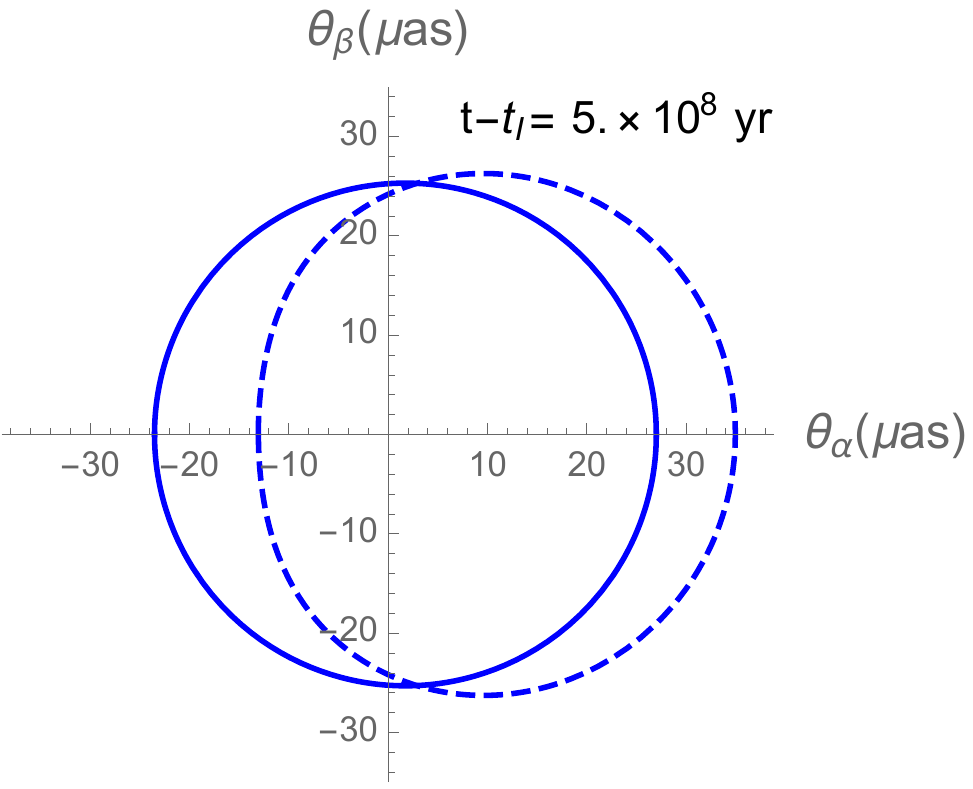}
\end{center}
\caption{\label{fig:ShadowevolutionSgrA*} 
Snapshots of the evolution of the shadow diameter, as seen by an equatorial observer, 
of a Sgr~A$^{\ast}$-type black hole with mass $M_0=4.2\times10^{6}M_{\odot}$ and $\alpha=0.05$. 
Exemplarily, we set the initial spin $\chi_{0}=0.95$ and scalar cloud mass $M_{c,0}=10^{-9}M_{0}$.
The top left plot corresponds to the beginning of phase~I, i.e., $t-t_{I}=0$. 
For comparison, we show the initial shadow diameter at the end of the superradiant evolution (bottom right) as a dashed line.
}
\end{figure*}

\begin{figure*}
\begin{center}
\includegraphics[width=.28\textwidth]{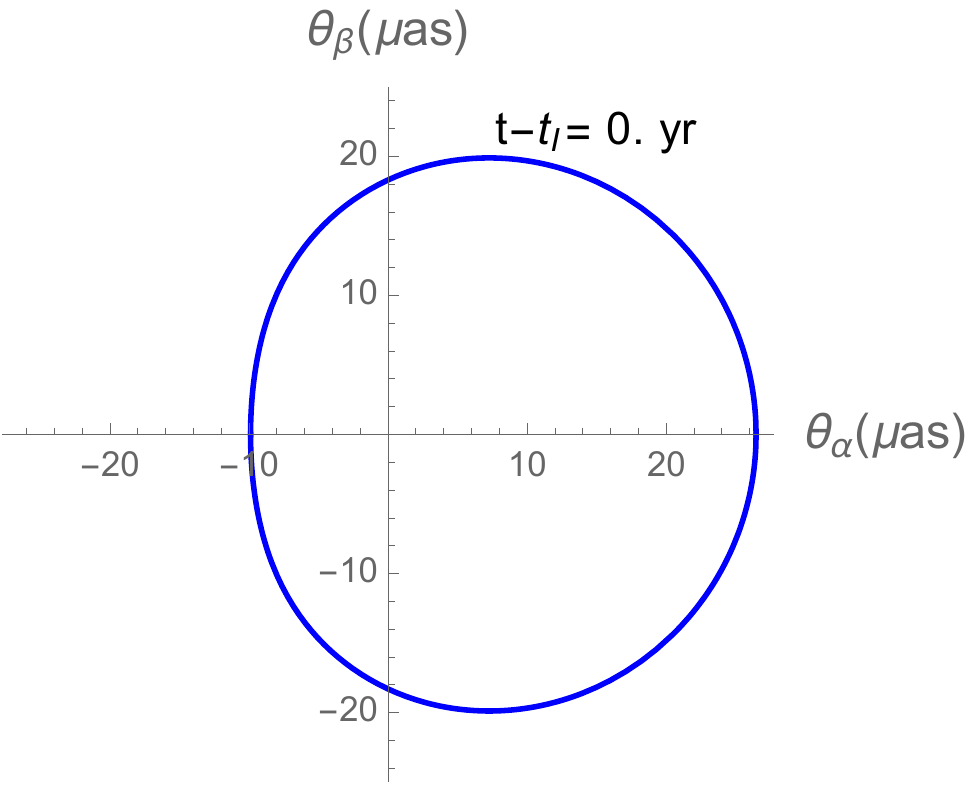}
\includegraphics[width=.28\textwidth]{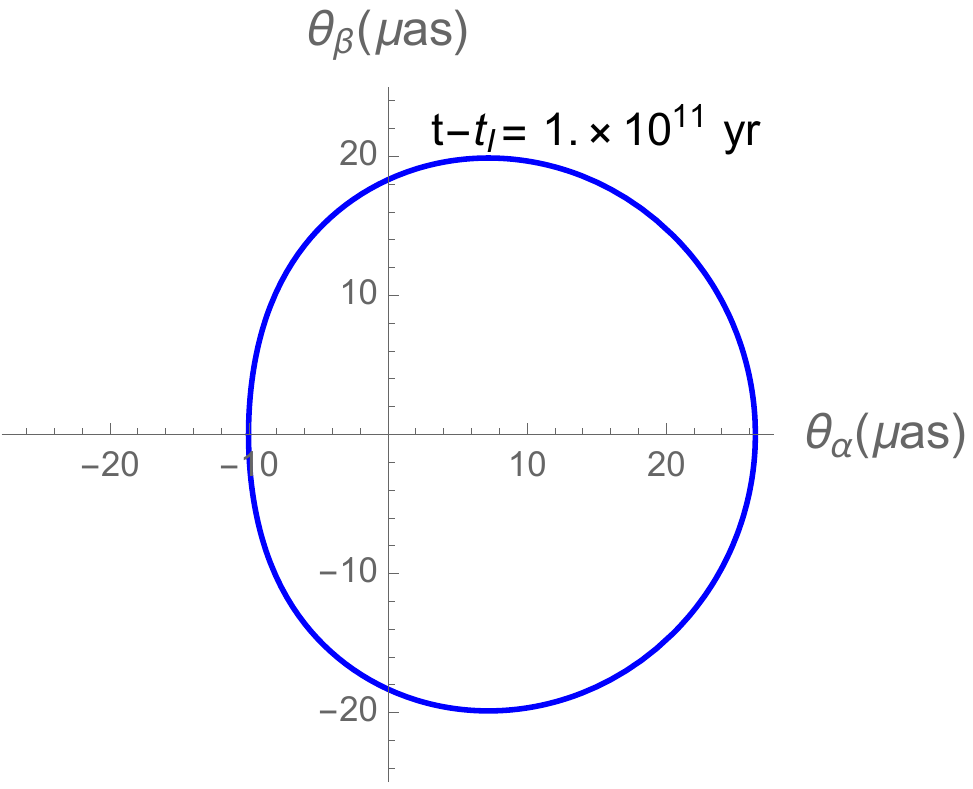}
\includegraphics[width=.28\textwidth]{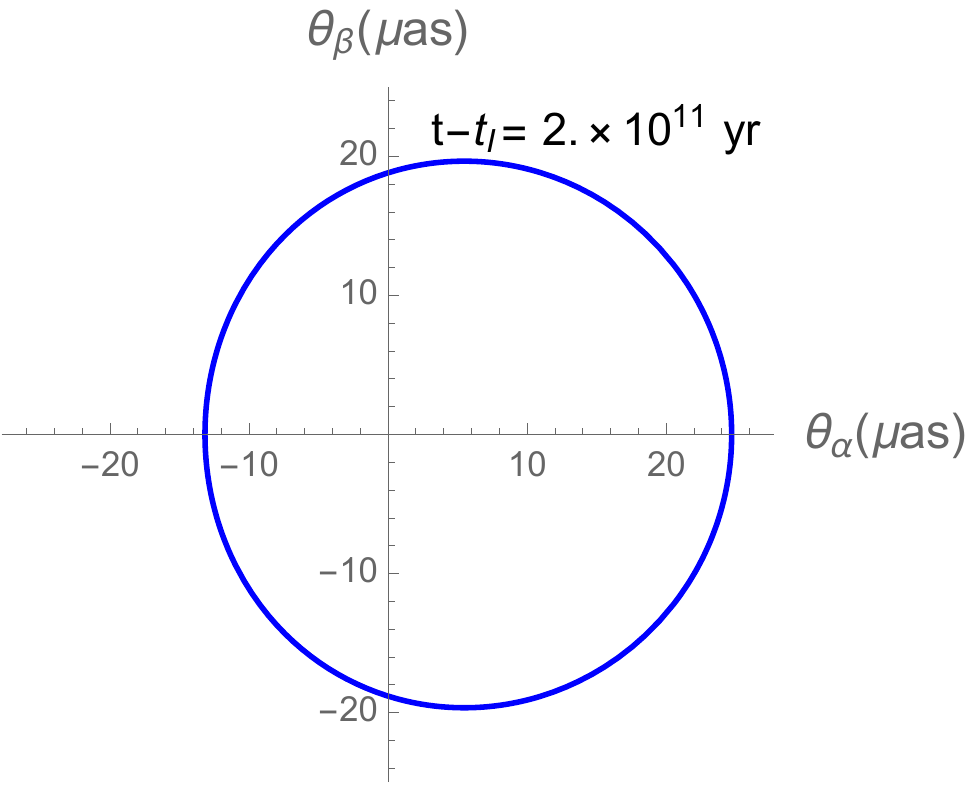}
\includegraphics[width=.28\textwidth]{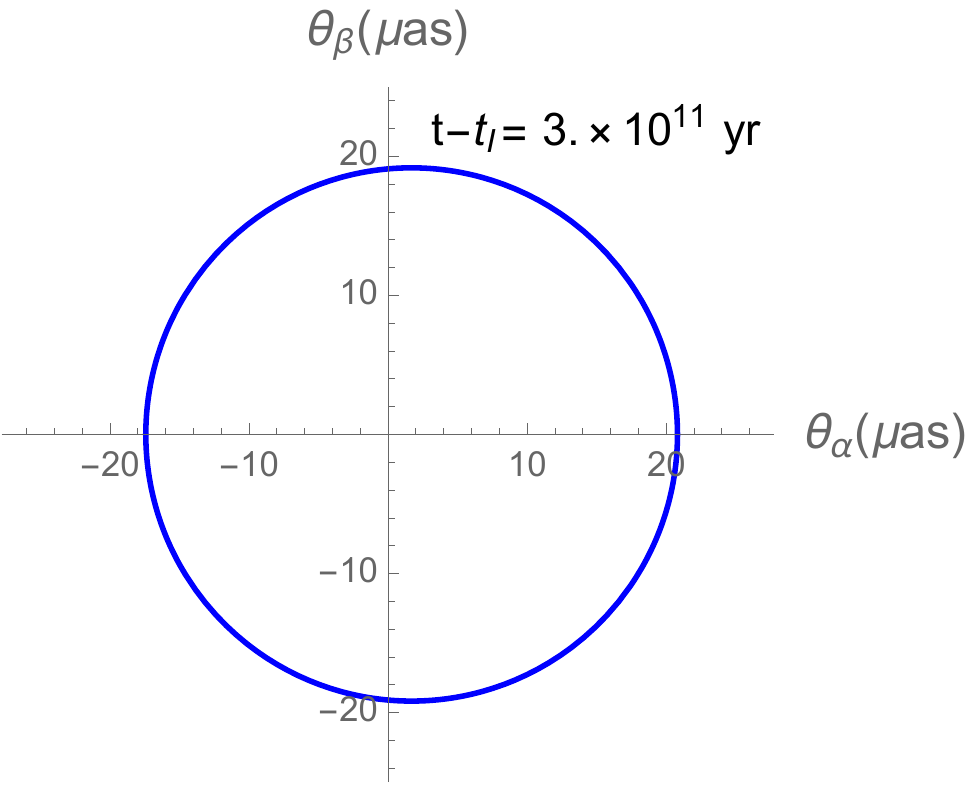}
\includegraphics[width=.28\textwidth]{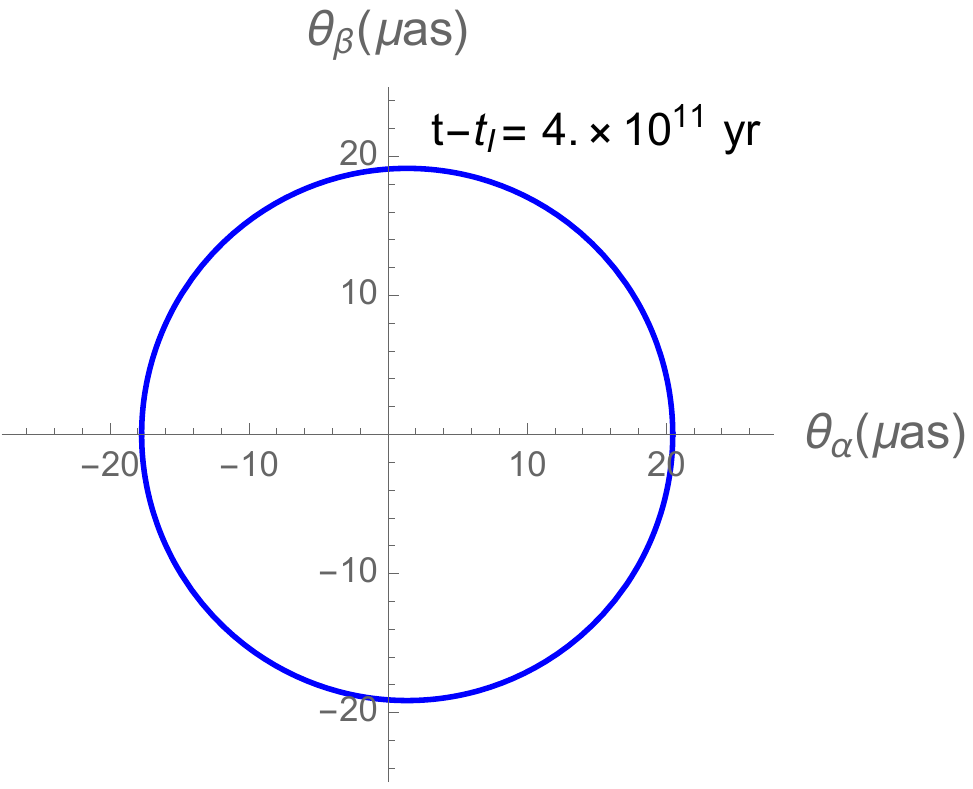}
\includegraphics[width=.28\textwidth]{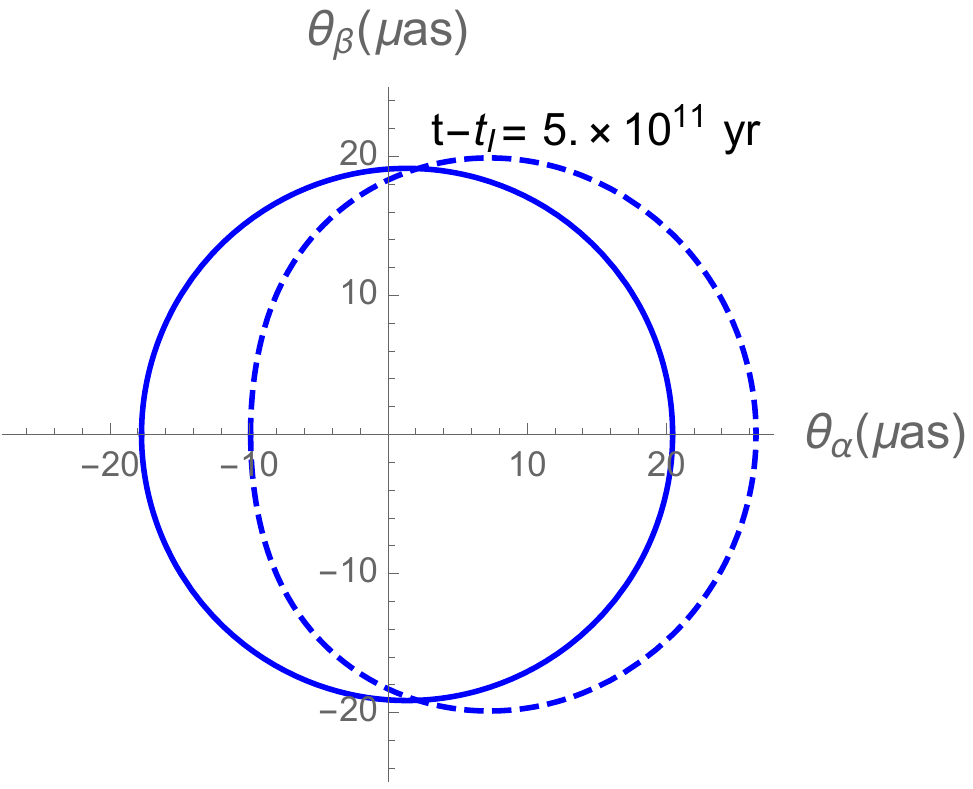}
\end{center}
\caption{\label{fig:ShadowevolutionM87} 
Same as Fig.~\ref{fig:ShadowevolutionSgrA*} but for a M87-like black hole with mass $M_0=6.5\times10^{9}M_{\odot}$.
}
\end{figure*}

We illustrate the evolution of the shadow diameter of Sgr~A$^{\ast}$ and M87
in Figs.~\ref{fig:ShadowevolutionSgrA*} and~\ref{fig:ShadowevolutionM87}, respectively.
In particular, we consider black holes with an initial spin of $\chi_{0}=0.95$ surrounded by scalar field fluctuations whose total initial mass is $10^{-9}M_{0}$ of the (initial) black hole mass.
We present snapshots of the shadow diameter at different stages of the entire superradiant evolution.
The last snapshot, corresponding to the end state of the evolution, superposes the initial (dashed lines) and final (solid lines) shadow.
We observe that during the superradiant evolution
the shadow diameters change by a few $\mu$as while their morphology changes from oblate due to high initial spin to a more spherical shape due to the low spin of the final black hole.
While this may give hope to observe a black hole undergoing a superradiant evolution,
we observe that the timescales of said evolution are $10^{8}\cdots10^{11}$yr.
So, how large is the change in the shadow diameter within a reasonable observation time of, say, a decade?
We addressed this question in the most promising regime, namely phase~I of the superradiant evolution where the black hole parameters and, hence, its shadow change most rapidly.
We summarize our results in Table~\ref{tab:tablesummaryBHexample}
for different values of the initial boson cloud.
Even in this most optimistic scenario, the change in the shadow diameter would be of $\mathcal{O}\left(10^{-6}~\mu\textrm{as}\right)$,
well below the sensitivity of current or future instruments. 
We remark that this conclusion is based on computation for fixed gravitational couplings $\alpha=0.05$
and may differ outside the small-coupling approximation. 
In the following section we explore more generally how the shadow evolution and its final state depends on the coupling.


\subsection{Shadow evolution parameter space}\label{ssec:ShadowParameterSpace}
So far we have studied the superradiance-driven evolution of the black hole shadow for a fixed value of the gravitational coupling, namely $\alpha=0.05$.
We observed that the final shadow (i.e., the shadow at the end of the superradiant evolution) is larger as compared to its initial value.
Although in Sec.~\ref{sec:shadowsuperradiance} we have discussed and justified this behavior, we left one question unanswered:
does the shadow diameter always increase?
We know that the (initial) parameters of the black hole and of the scalar condensate play a crucial role in the evolution,
and so determine the shape of the shadow.

\begin{figure}[htpb!]
\begin{center}
\includegraphics[width=0.45\textwidth,clip]{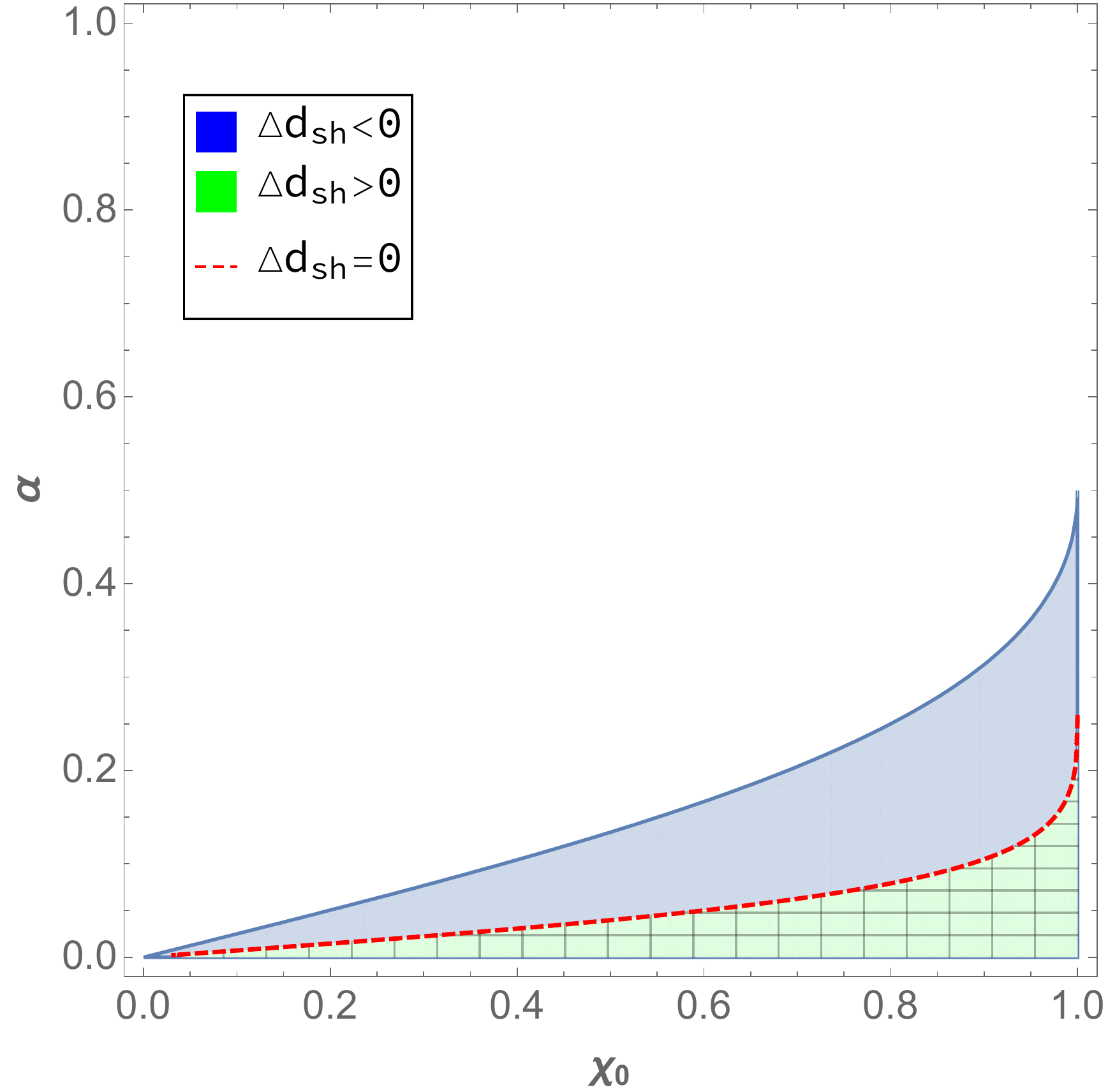}
\caption{\label{fig:Deltashsignparameterspace}
Dependence of the sign of $\Delta{d_{\rm sh}}$ on the gravitational coupling $\alpha$ and the initial spin $\chiZ$
for the $l=m=1$ mode.
The blank space corresponds to parameters that do not satisfy the superradiance condition.
For $m=l>1$ the behavior is qualitatively similar.
}
\end{center}
\end{figure}

In order to study the end state of the black hole shadow, and its dependence on the initial configurations,
let us define the change of the angular diameter $\Delta d_{\rm sh}=d_{\rm sh,II} - d_{\rm sh,0}$,
where $d_{\rm sh,II}$ denotes the final state in phase~II and $ d_{\rm sh,0}$ the initial one; see Sec.~\ref{sec:Evolution}.
Using Eq.~\eqref{eq:shadowdiameteranalytical} the change in the shadow is determined by
\begin{align}
\label{eq:DeltaShadow}
\Delta d_{\rm sh} = & \frac{9}{r_o}\left(\MII-\MZ \right)
\\ &
        + \frac{3(2\sqrt{3} - 3)}{r_o} \left[\MII (1-\chiII^2)^{\delta} - \MZ (1-\chiZ^2)^{\delta} \right]
\,.\nonumber
\end{align}
This clearly indicates that the change in the shadow can be negative, zero, or positive.
The specific case, $\Delta{d_{\rm sh}}\lesseqqgtr0$, is determined by the condition
\begin{align}
\label{eq:Shadowsigncondition}
\frac{\MII(1-\chiII^2)^\delta-M_0(1-\chi_0^2)^\delta}{M_0-\MII} \lesseqqgtr & \frac{3}{2\sqrt{3}-3}
\,,
\end{align}
as follows from Eq.~\eqref{eq:DeltaShadow}.
We can relate the condition to the initial black hole parameters $(\MZ,\chiZ)$ and the gravitational coupling
by using Eqs.~\eqref{eq:finalmassBH} and~\eqref{eq:finalBHspin}.
The results are shown in Fig.~\ref{fig:Deltashsignparameterspace} in which we present the signature of $\Delta d_{\rm sh}$ in the 
phase space spanned by the gravitational coupling $\alpha$ and initial black hole spin $\chiZ$.
Here we impose both Eq.~\eqref{eq:Shadowsigncondition} and the superradiance condition~\eqref{eq:sprcond} rewritten as
\begin{align}
\label{eq:alphaSRcond}
\alpha \leq &\frac{m}{2}\frac{\chi_0}{1+\sqrt{1-\chi_0^2}}
\,,
\end{align}
where we set $\omega_{\rm R}\sim \mu$.
The green, rectangular-patterned region corresponds to $\Delta{d_{\rm sh}}>0$, i.e. a final shadow larger than the initial one.
The blue region corresponds to $\Delta{d_{\rm sh}}<0$, where the final shadow is smaller than the shadow at the beginning of the superradiant evolution. 
The red dashed curve indicates the separatrix for which $\Delta{d_{\rm sh}}=0$.
As we see, by fixing $\alpha=0.05$ and looking at large initial spins, we were restricting ourselves to the region in which $\Delta{d_{\rm sh}}>0$.
Looking at higher values of the gravitational coupling $\alpha\sim\mathcal{O}(0.1)$, this is no longer the case, and the final shadow can indeed be smaller than the initial one after the superradiant evolution.
This is not surprising: as indicated in Eq.~\eqref{eq:dshEvolPean} the change of the shadow is determined by
$\frac{\dif d_{\rm sh}}{\dif M} + \frac{m\,\MZ}{\alpha} \frac{\dif d_{\rm sh}}{\dif J}$,
and these two terms have opposite signs so lead to competing effects. 

\begin{figure}[h!]
\begin{center}
\includegraphics[width=0.45\textwidth,clip]{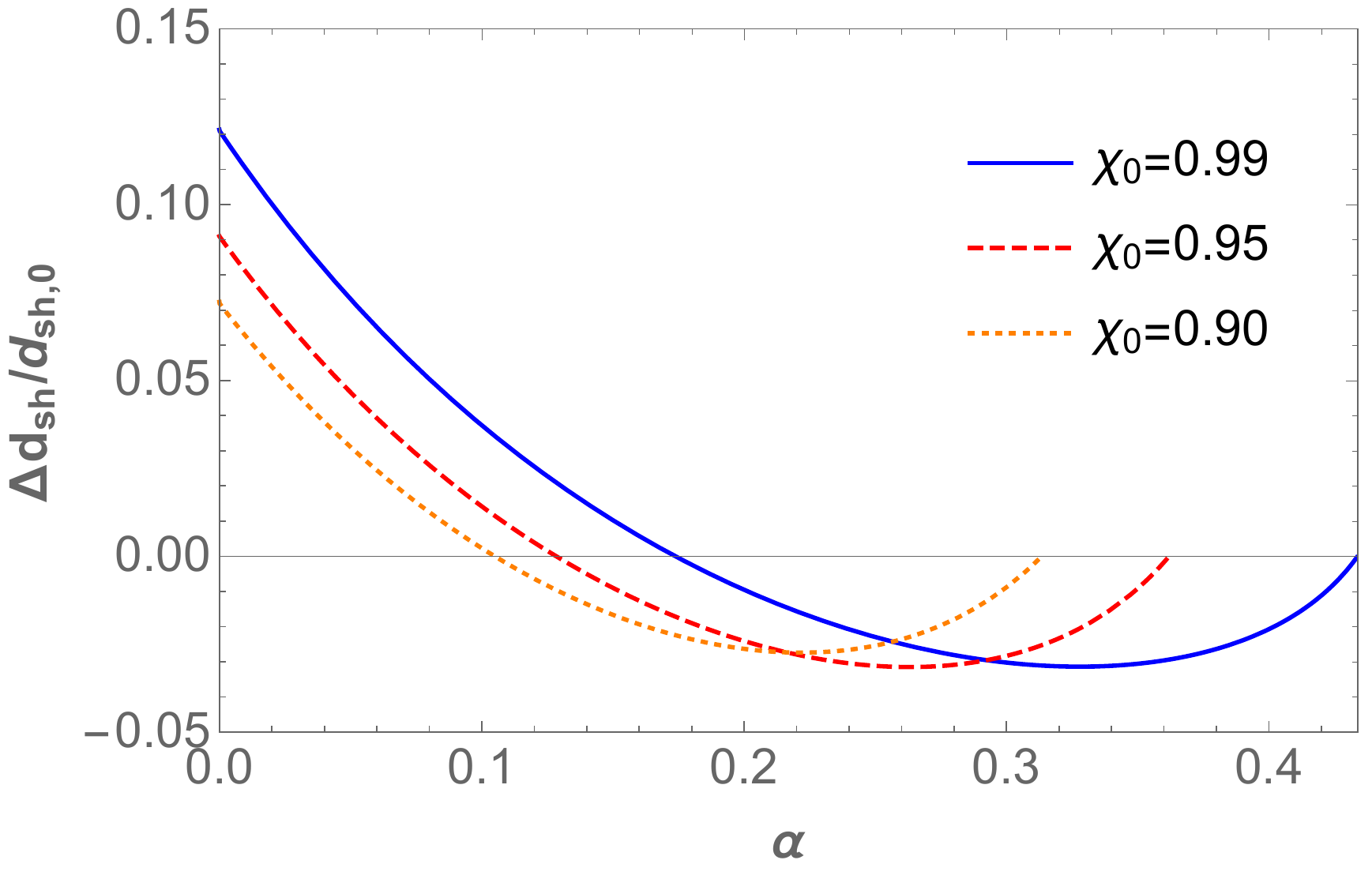}
\caption{\label{fig:Deltashadowrelalphal1}
Relative change of the final shadow with respect to the initial shadow as a function of the gravitational 
coupling constant $\alpha$ for different values of the initial dimensionless spin $\chi_0$. 
We have fixed $m=l=1$. For $m=l>1$ the behavior is qualitatively similar. 
}
\end{center}
\end{figure}

To quantify this effect, in Fig.~\ref{fig:Deltashadowrelalphal1} we show the relative change $\Delta{d_{\rm sh}}/d_{\rm sh,0}$ 
as a function of the gravitational coupling constant for different spins. 
We observe that this relative difference increases for decreasing coupling $\alpha$ and, in the small-coupling regime, for increasing the initial spin of the black hole.
Additionally, we verify that the zero crossings in Fig.~\ref{fig:Deltashadowrelalphal1} correspond to the $\Delta d_{\rm sh}=0$ lines in Fig.~\ref{fig:Deltashsignparameterspace}.
Here we present the calculations for $l=m=1$, but we verified that the behavior for $m=l>1$ is qualitatively similar.
We remark, however, that the values of $\alpha$ for which the system is in the superradiant regime are directly proportional to the mode number $m$; cf. Eq.~\eqref{eq:alphaSRcond}.

Together with Fig.~\ref{fig:Deltashadowrelalphal1} we can study the maximum and minimum values of $\Delta{d_{\rm sh}}/d_{\rm sh,0}$. 
Let us focus first on the maximum change: this is reached when all angular momentum is extracted and the final black hole is a Schwarzschild black hole.
We reach this state in the limit that $\alpha\to0$.
Then, the final black hole has $\chiII=0$ and $\MII=M_0$, and the change of the shadow is bounded by
\begin{align}
\frac{\Delta{d_{\rm sh}}^{\rm max}}{d_{\rm sh,0}}=\frac{(2\sqrt{3}-3)\left[1-(1-\chi_{0}^2)^\delta\right]}{3+(2\sqrt{3}-3)(1-\chi_{0}^2)^\delta}
\,.
\end{align}
Note that this maximum depends only on the initial dimensionless spin $\chi_0$. 

Now let us determine the minimum.
Therefore, we compute $d(\Delta{d_{\rm sh}}/d_{\rm sh,0})/d\alpha=0$, which translates into the condition $d\MII/d\alpha=0$, where $\MII$ is given by Eq.~\eqref{eq:finalmassBH}. The latter condition yields a fifth-order polynomial in $\alpha$ and is not possible to solve analytically. In order to solve for the value of the gravitational coupling constant that minimizes $\Delta{d_{\rm sh}}/d_{\rm sh,0}$, $\alpha_{\rm min}$, we computed the minimum numerically for different modes. We observed that the value of $\alpha_{\rm min}$ for $m=l>1$ is directly proportional to the value for $m=l=1$. Therefore, $\alpha_{\rm min}=\tilde{\alpha}m$, where $\tilde{\alpha}$ is the minimum computed for $m=l=1$. Substituting $\alpha_{\rm min}$ in Eq.~\eqref{eq:finalBHspin} and Eq.~\eqref{eq:finalmassBH} gives
\begin{align}
\tilde{\chi}_{\rm II}\equiv\chiII(\alpha=\alpha_{\rm min})&=\frac{4\tilde{\alpha}}{1+4\tilde{\alpha}^2}~,\\
{\tilde{M}_{\rm II}}\equiv\MII(\alpha=\alpha_{\rm min})&=M_0\frac{1-\sqrt{1-16\tilde{\alpha}^2(1-\tilde{\alpha}\chi_{0})^2}}{8(1-\tilde{\alpha}\chi_{0})}~.
\end{align}
Since the minimum of $\Delta{d_{\rm sh}}/d_{\rm sh,0}$ depends only on $\chiII$ and $\MII$ [see Eq.~\eqref{eq:shadowdiameteranalytical}], it will be independent of the mode $m$. This value is given by
\begin{align}
\frac{\Delta{d_{\rm sh}}^{\rm min}}{d_{\rm sh,0}}=&\frac{3(\frac{{\tilde{M}_{\rm II}}}{M_0}-1)}{3+(2\sqrt{3}-3)(1-\chi_0^2)^\delta}\nonumber\\&+\frac{(2\sqrt{3}-3)\left[\frac{{\tilde{M}_{\rm II}}}{M_0}(1-\tilde{\chi}_{\rm II}^2)^\delta-(1-\chi_0^2)^\delta\right]}{3+(2\sqrt{3}-3)(1-\chi_0^2)^\delta}~.
\end{align}
Therefore, for a given initial spin $\chi_0$, the relative change in the shadow will be bounded by
\begin{align}
\frac{\Delta{d_{\rm sh}}^{\rm max}}{d_{\rm sh,0}}>\frac{\Delta{d_{\rm sh}}}{d_{\rm sh,0}}\geq\frac{\Delta{d_{\rm sh}}^{\rm min}}{d_{\rm sh,0}}~,
\end{align}
independently of the mode $m$. 
For example, if $\chi_0=0.99$ we obtain $\Delta{d_{\rm sh}}^{\rm max}/d_{\rm sh,0}=12.1\times10^{-2}$ and $\Delta{d_{\rm sh}}^{\rm min}/d_{\rm sh,0}=-3.1\times10^{-2}$.

\section{Conclusions and outlook}
Black holes have become unique gravitational probes for ultralight,
beyond--standard model particles, including fashionable dark matter candidates or the string axiverse~\cite{Arvanitaki:2010sy,Brito:2015oca,Hui:2016ltb,Baumann:2018vus}.
The underlying phenomenon is black hole superradiance, i.e., a classical mechanism that leads to the buildup of bosonic condensates that are composed of
low-frequency, ultralight fields.

The majority of studies in this context focuses on the detectability of these clouds with gravitational waves.
We, instead, placed observations of the black hole shadow into the spotlight.
While Refs.~\cite{Cunha:2015yba,Cunha:2019ikd,Roy:2019esk}
explored the modification of the shadow due to the presence of a bosonic cloud,
we investigated the impact of the superradiant evolution on the black hole shadow in a wide range of parameter space.

To do so, we developed a numerical code capable of simulating the buildup of the gravitational atom in the adiabatic approximation
(following Refs.~\cite{Brito:2014wla,Ficarra:2018rfu})
and simulating the resulting evolution of the shadow diameter.
We have complemented this numerical study with analytic approximations to efficiently model the superradiant evolution. 
Our techniques are valid for any black hole mass, although the presentation focuses on M87$^{\ast}$ and Sgr~A$^{\ast}$, 
the black holes at the center of the galaxy M87 and of the Milky Way that are prime candidates for the EHT.

The superradiant evolution exerts two competing effects on the black hole shadow:
\begin{enumerate*}[label={(\roman*)}]
\item the decrease of the black hole mass {\textit{decreases}} the shadow diameter;
while
\item the decrease of the black hole spin {\textit{increases}} the shadow diameter. 
\end{enumerate*}
The majority of our study was performed in the small coupling regime, where $\alpha\ll1$ and the adiabatic approximation is valid.
In this regime, the spin effects appear to dominate and the black hole shadow increases over time. These changes can be as large as a few $\mu$as
as is illustrated in a series of snapshots in Figs.~\ref{fig:ShadowevolutionSgrA*} and~\ref{fig:ShadowevolutionM87}.
However, for supermassive black holes and for $\alpha\ll1$ this change occurs over timescales of $10^{8}\cdots10^{11}$ yr. That is,
in practice these effects will not be detectable with VLBIs over realistic observation times.

For large couplings $\alpha\sim\mathcal{O}(0.1)$ 
however, the evolution of the black hole mass seems to dominate and the black hole shadow diameter would decrease.
To estimate the involved superradiance timescales, take
the fastest growth rate of $M\,\Gamma = 1.5\times10^{-7}$
found for couplings of $\alpha=0.42$ and spins of $\chi=0.99$~\cite{Dolan:2007mj}.
That is, the shortest possible 
timescale is $\tau\sim7\times10^{6}M\sim 420\left(\frac{M}{M_{\odot}}\right)$s,
or $\tau\sim9\times10^{4}$yr for a
black hole of $M\sim6.5\times10^{9}M_{\odot}$.
Indeed, a closer qualitative inspection revealed that sufficiently small (large) gravitational couplings yield a decrease (increase)
of the shadow diameter directly induced by the superradiant evolution; see Fig.~\ref{fig:Deltashsignparameterspace}.

Although our original question ``Can we tape the superradiant evolution with observations of the black hole shadow?''
has to be negated, this project has been very instructive:
It has taught us the richness of effects of the superradiant evolution on the black hole shadow which is significantly more complex 
than initially expected.
In particular, it is not a clear-cut, one-fits-all observable as was concluded in Ref.~\cite{Roy:2019esk}.
Furthermore, although the overall change in the shadow diameter can be a few $\mu$as it has to be compared to the actual observation time. Even if we assume that
the EHT, or a follow-up project, would observe the shadow evolution over several decades, the change during that time is several orders of magnitude
below their resolution.

The present paper has focused solely on the superradiant evolution to cleanly identify its impact on the evolution of the black hole and its shadow.
We neglected additional phenomena such as accretion of ordinary matter that would have the opposite effect.
For simplicity, we kept the distance between observer and black hole constant. Given the cosmological timescales involved, 
it would be interesting to include the cosmological evolution of the black hole's distance to us.
We leave a detailed analysis of these effects for future work.

\begin{acknowledgments}
We thank P.~Jonker and H.~Okada da~Silva for useful discussions.
H.W. acknowledges financial support by the Royal Society, UK, via her Royal Society University Research Fellowship, Grant No. UF160547,
and the Royal Society Research Grant No. RGF$\backslash$R1$\backslash$180073.
We thankfully acknowledge the computer resources and the technical support provided by the 
Leibniz Supercomputing Center via PRACE Grant No. 2018194669 ``FunPhysGW: Fundamental Physics in the era of gravitational waves''
and by the DiRAC Consortium via 
STFC DiRAC Grants No. ACTP186, No. ACDP191 and No. ACSP218.
\end{acknowledgments}

\appendix
\section{Fitting formulas for superradiant evolution}\label{app:BHspinevolutionapp}
In this appendix we derive the fitting formulas presented in Sec.~\ref{ssec:modelizingsuperradiance}. 
They will allow us to approximate the superradiant evolution analytically.
Since the improved gamma fit is analogous to the gamma fit (but with a different integral to solve) we will derive the former in less detail than the gamma fit and squared fit.

\subsection{Squared fit}
In this fit we develop a fitting formula that uses an ansatz for the black hole spin. Noticing the exponential behavior of the numerical solution, our ansatz consists of
an exponential function with a second order time dependence. This is because for this case the change is fast enough such that we have to take into account 2 orders in the time evolution. 
This scheme appears well suited to model large seeds, whereas the evolution of small seeds is not well captured as is illustrated in Fig.~\ref{fig:SpinevolutionM87}.
Specifically, the ansatz is given by
\begin{align}
\label{appeq:AnsatzSquaredFit}
J(t) = & A e^{-\gamma{t} - \beta{t^2}}+B
\,.
\end{align}

Applying the conditions $J(0)\equiv{J_0}$ and $J(t\rightarrow\infty)=J_{\rm II}$ determines the
coefficients
\begin{align*}
A = & \left(J_0-J_{\rm II}\right)
\,,\qquad 
B=J_{\rm II}
\,.
\end{align*}
In order to compute the exponent $\gamma$ we will use the spin differential equation
\begin{align}
\label{appeq:JdotVsMc}
\frac{dJ}{dt} = & -\frac{2 m}{\mu}\Gamma_{0} M_c
\,,
\end{align}
where $\Gamma_{0}$ is the (initial) growth or decay rate given in Eq.~\eqref{eq:SFFreqIm}.\footnote{Notice that we suppress the subscripts ``nlm'' to improve readability}.
Evaluating this relation at $t=0$ and using the relation $\dif J = \frac{m}{\mu}\dif M$, we find the coefficient
\begin{align}
\label{appeq:SquaredFitgamma}
\gamma = & \frac{2m\,\Gamma_{0}}{\mu}\frac{M_{c,0}}{J_0-J_{\rm II}}
=   2 \Gamma_{0} \frac{M_{c,0}}{M_0-\MII}
\,,
\end{align}
where $\MII$ is given by Eq.~\eqref{eq:finalmassBH}.

In order to compute the coefficient $\beta$ in ansatz~\eqref{appeq:AnsatzSquaredFit}
we need an extra condition.
For this purpose we consider the mean value of the evolved quantities, denoted by 
\begin{align}
J_{\ast} \equiv J(t=t_{\ast}) = & \frac{J_{0} + J_{\rm II}}{2}
\,,
\end{align}
and likewise for all other variables. We denote the time when the mean values are reached as $t_{\ast}$.
We can estimate $t_{\ast}$ by using Eq.~\eqref{appeq:JdotVsMc}, i.e.,
\begin{align}
\left.\frac{\dif J}{\dif t} \right|_{t=t_{\ast}} = & 
-\frac{2 m}{\mu} \Gamma_{\ast}\,M_{c,\ast}
\sim \frac{J_{\ast} - J_{0}}{t_{\ast}}
\,,
\end{align}
where $\Gamma_{\ast}=\frac{\Gamma_{0}}{2}$,
to find
\begin{align}
t_{\ast} = & \frac{\mu}{m\Gamma_{0}} \frac{J_{0}-J_{\rm II}}{M_{c,0}+M_{c,\rm II}}
=   \frac{2}{\gamma} \frac{M_{c,0}}{M_{c,0}+M_{c,\rm II}}
\,.
\end{align}
Ansatz~\eqref{appeq:AnsatzSquaredFit}
now gives the extra condition
\begin{align*}
J_{\ast} = & \frac{J_{0}+J_{\rm II}}{2}
= J_{\rm II} + \left(J_{0}-J_{\rm II}\right) \exp\left[-\gamma t_{\ast} - \beta t^{2}_{\ast} \right]
\,,
\end{align*}
that we solve to find
\begin{align}
\label{appeq:SquaredFitbeta}
\beta = & \frac{\ln\,2}{t^2_{\ast}} - \frac{\gamma}{t_{\ast}}
\,.
\end{align}
Finally, the fitting formula for the black hole spin is
\begin{align}
\label{appeq:SquaredFitFormulaSpin}
J(t) = & J_{\rm II} + \left(J_{0}-J_{\rm II}\right) \exp\left[-\gamma t - \beta t^{2} \right]
\,,
\end{align}
with the exponents $\beta$ and $\gamma$ given in Eqs.~\eqref{appeq:SquaredFitbeta} and~\eqref{appeq:SquaredFitgamma}.
One can now repeat the same procedure for the mass $M$ of the black hole as well as the mass $M_{c}$ and spin $J_{c}$ of the cloud.
In general, the fitting formula for the parameter $p(t)$ reads
\begin{align}
p(t)=(p_0-p_{\rm II})e^{-\gamma{t}-\beta{t^2}}+p_{\rm II}~,
\end{align}
with
\begin{align}
\label{appeq:SquaredFitGeneralgamma}
\gamma=\left.\frac{dp}{dt}\right|_{t=0}\frac{1}{p_0-p_{\rm II}},\quad\beta=\frac{\ln(2)}{t_{*}^2}-\frac{\gamma}{t_{*}}~.
\end{align}

\subsection{Gamma fit}
In this fit, instead of directly modeling the black hole parameters we model the time dependence of the imaginary part of the frequency. This is, as the black hole parameters change due to the superradiant evolution, so will the decay
or growth rate of the bosonic field determined by the parameters.
We capture this time dependence of the imaginary part of the frequency
with the ansatz
\begin{align}
\label{appeq:AnsatzGammaFit}
\Gamma(t) = & \Gamma_{0} \exp\left[-\gamma t\right]
\,,
\end{align}
where we suppressed the subscript ``(nlm)'' for readability and $\Gamma_{0}$ is the
rate given in Eq.~\eqref{eq:SFFreqIm}.
Substituting this ansatz into the evolution equation for the cloud's mass, Eq.~\eqref{eq:energyinstabilityflux}, yields
\begin{align}
\label{appeq:GammaFitODEMc}
\frac{\dif M_c}{\dif t} = & 2\Gamma_{0} M_c \exp\left[-\gamma{t}\right]
\,.
\end{align}
Solving the differential equation, we obtain
\begin{align}
\label{appeq:ansatzMcapp}
M_{c} (t) = & M_{c,0}\exp\left[\frac{2\Gamma_{0}}{\gamma}\left(1-e^{-\gamma{t}}\right)\right]
\,,
\end{align}
where we imposed $\lim\limits_{t\rightarrow0} M_{c} = M_{c,0}$.
We can read off the exponent $\gamma$ by considering the limit
$\lim\limits_{t\rightarrow{\infty}}M_c(t)=M_{c,\rm II}$, and find
\begin{align}
\label{appeq:GammaFitExponent}
\gamma = & \frac{2\Gamma_0}{\ln\left(\frac{M_{c,\rm II}}{M_{c,0}}\right)}
\,.
\end{align}
Substituting Eq.~\eqref{appeq:ansatzMcapp}
into the evolution equations~\eqref{eq:spinevolutionsdiffeqs}, we obtain
\begin{subequations}
\label{appeq:GammaFitspinevolutionsdiffeqs}
\begin{align}
\frac{dM}{dt}   = & - \frac{dM_c}{dt}
\\ = &
        - 2 \Gamma_{0} M_{c,0}\, e^{-\gamma t}
        \exp\left[\frac{2\Gamma_{0}}{\gamma} \left(1-e^{-\gamma t}\right) \right]
\nonumber\,, \\ 
\frac{dJ}{dt}   = & - \frac{m}{\mu}\frac{dM_c}{dt}
\\ = &
        - 2 \frac{m}{\mu} \Gamma_{0} M_{c,0}\, e^{-\gamma t}
        \exp\left[\frac{2\Gamma_{0}}{\gamma} \left(1-e^{-\gamma t}\right) \right]
\,. \nonumber
\end{align}
\end{subequations}
We can integrate these differential equations to
\begin{subequations}
\begin{align}
M(t) = & M_{0} - \left(M_{c}(t) - M_{c,0} \right)
\\ = &
        M_{0} 
        - M_{c,0} \left\{\exp\left[\frac{2\Gamma_{0}}{\gamma} \left(1-e^{-\gamma\,t}\right) \right] - 1 \right\}
\,,\nonumber\\
J(t) = & J_{0} - \frac{m}{\mu} \left( M_{c}(t)- M_{c,0} \right)
\\ = &
        J_{0}
        - M_{c,0} \frac{m}{\mu} \left\{\exp\left[\frac{2\Gamma_{0}}{\gamma} \left(1-e^{-\gamma\,t}\right) \right] - 1 \right\}
\,,\nonumber
\end{align}
\end{subequations}

\subsection{Improved gamma fit}
We improve our model~\eqref{appeq:AnsatzGammaFit}
by taking the ansatz
\begin{align}
\label{appeq:AnsatzImprovedGammaFit}
\Gamma(t) = & \Gamma_{0} \exp\left[1-e^{\gamma t}\right]
\,,
\end{align}
and call this {\textit{improved gamma fit}}.
Then, the evolution of the cloud's mass is determined by
\begin{align}
\label{appeq:GammaFitODEMcimpr}
\frac{\dif M_c}{\dif t} = & 2\Gamma_{0} M_c \exp\left[1-e^{\gamma t}\right]
\,.
\end{align}
We integrate it to find
\begin{align}
M_c(t) = & M_{c,0}\exp\left\{2\Gamma_{0}\int_{0}^{t}\exp\left(1-e^{\gamma t'}\right)dt'\right\}
\nonumber\\ = & 
        M_{c,0}\exp\left\{2\Gamma_{0}\left[\frac{e}{\gamma}E[-e^{\gamma{t'}}]\right]_0^t\right\}
\nonumber\\ = & 
        M_{c,0}\exp\left\{\frac{2\Gamma_{0}e}{\gamma}\left(E[-e^{\gamma{t}}]-E[-1]\right)\right\}
\,,
\end{align}
where $E[x]$ is the exponential integral defined as
\begin{align*}
E[x]
= & -\int_{-x}^{\infty}\frac{e^{-t}}{t}dt
\,.
\end{align*}
In particular, $E[-1]=-0.219384$,  $E[-\infty]=0$. 
Proceeding analogously to the previous case, the spin and mass are given by
\begin{subequations}
\begin{align}
\label{appeq:ansatzJimpapp}
J(t) = & J_0-\frac{m}{\mu}M_{c,0}\left\{\exp\left[\frac{2\Gamma_0e}{\gamma}\left(E[-e^{\gamma{t}}]-E[-1]\right)\right]-1\right\}
\,,\\
\label{appeq:ansatzMimpapp}
M(t) = & M_0-M_{c,0}\left\{\exp\left[\frac{2\Gamma_0e}{\gamma}\left(E[-e^{\gamma{t}}]-E[-1]\right)\right]-1\right\}
\,,
\end{align}
\end{subequations}
with 
\begin{align}\label{appeq:improgammaapp}
\gamma = & -\frac{2\Gamma_0E[-1]e}{\ln\left({\frac{M_{c,\rm II}}{M_{c,0}}}\right)}
\,.
\end{align}
As we show in Fig.~\ref{fig:SpinevolutionM87}
in the main text, the improved gamma fit provides good results for both small and large seeds of the bosonic cloud.

\subsection{Heaviside tuning fit}
For this fit we will model the time dependence of the imaginary part of the frequency using an analytic approximation to the Heaviside step function $\Theta(t)$,
\begin{align}
\Theta(t)=\lim_{\gamma\rightarrow\infty}\frac{1}{1+e^{\gamma{t}}}~,
\end{align}
where $\gamma$ is a coefficient that regulates how close the  the analytic approximation is to Heaviside step function. Using this expression we write
\begin{align}
\label{appeq:AnsatzImprovedGammaFit}
\Gamma(t) = & \Gamma_{0} \frac{2}{1+e^{\gamma{t}}}
\,.
\end{align}
Proceeding in the same way as the other fits we have to solve the equation
\begin{align}
\label{appeq:GammaFitODEMcimpr}
\frac{\dif M_c}{\dif t} = & 4\Gamma_{0} M_c \frac{1}{1+e^{\gamma{t}}}
\,,
\end{align}
which yields
\begin{align}\label{appeq:McHeaviside}
M_c(t)=M_{c,0}\left(\frac{1+e^{-\gamma{t}}}{2}\right)^{\frac{4\Gamma_{0}}{\gamma}}~,
\end{align}
with
\begin{align}
\gamma=4\Gamma_{0}\frac{\ln(2)}{\ln\left(\frac{M_{c,0}}{M_{c,\rm II}}\right)}~.
\end{align}
We find that for the large seed this fit performs better than the others. For the small seed case this fit is not good enough. If we focus on the small seed case we can modify expression \eqref{appeq:McHeaviside} substituting $t\rightarrow{t-t_0}$ and $\gamma\rightarrow{\gamma/k}$ on the exponential. The first change will make the slope of the step higher while the second one will shift the position of the step. By comparison with numerical data we find that the best choices for this tuning are
\begin{align}
t_0=5.5\tau_{SR},\qquad k=1/7~.
\end{align}
Therefore, the final expression for $M_c$ reads
\begin{align}\label{appeq:McHeavisidemod}
M_c(t)=M_{c,0}\left(\frac{1+e^{-\frac{\gamma}{k}{(t-t_0)}}}{2}\right)^{\frac{4\Gamma_{0}}{\gamma}}~,
\end{align}
with $\{t_0=0,k=1\}$ for the large seed and $\{t_0=5.5\tau_{SR},k=1/7\}$ for the small seed. Although this modification can be done to the other fits, this is the one that adjusts better to the numerical simulations. The mass and spin of the black hole read
\begin{subequations}
	\begin{align}
	M(t) = & M_{0} - \left(M_{c}(t) - M_{c}(0) \right)
	\\ = &
	M_{0} 
	- \nonumber\\& M_{c,0} \left[\left(\frac{1+e^{-\frac{\gamma}{k}{(t-t_0)}}}{2}\right)^{\frac{4\Gamma_{0}}{\gamma}}-\left(\frac{1+e^{\frac{\gamma}{k}{t_0}}}{2}\right)^{\frac{4\Gamma_{0}}{\gamma}} \right]
	\,,\nonumber\\
	J(t) = & J_{0} - \frac{m}{\mu} \left( M_{c}(t) - M_{c,0} \right)
	\\ = &
	J_{0}
	-  \nonumber\\&\frac{m}{\mu}M_{c,0} \left[\left(\frac{1+e^{-\frac{\gamma}{k}{(t-t_0)}}}{2}\right)^{\frac{4\Gamma_{0}}{\gamma}}-\left(\frac{1+e^{\frac{\gamma}{k}{t_0}}}{2}\right)^{\frac{4\Gamma_{0}}{\gamma}} \right]
	\,,\nonumber
	\end{align}
\end{subequations}

\begin{figure}[h!]
	\begin{center}
		\includegraphics[width=0.475\textwidth,clip]{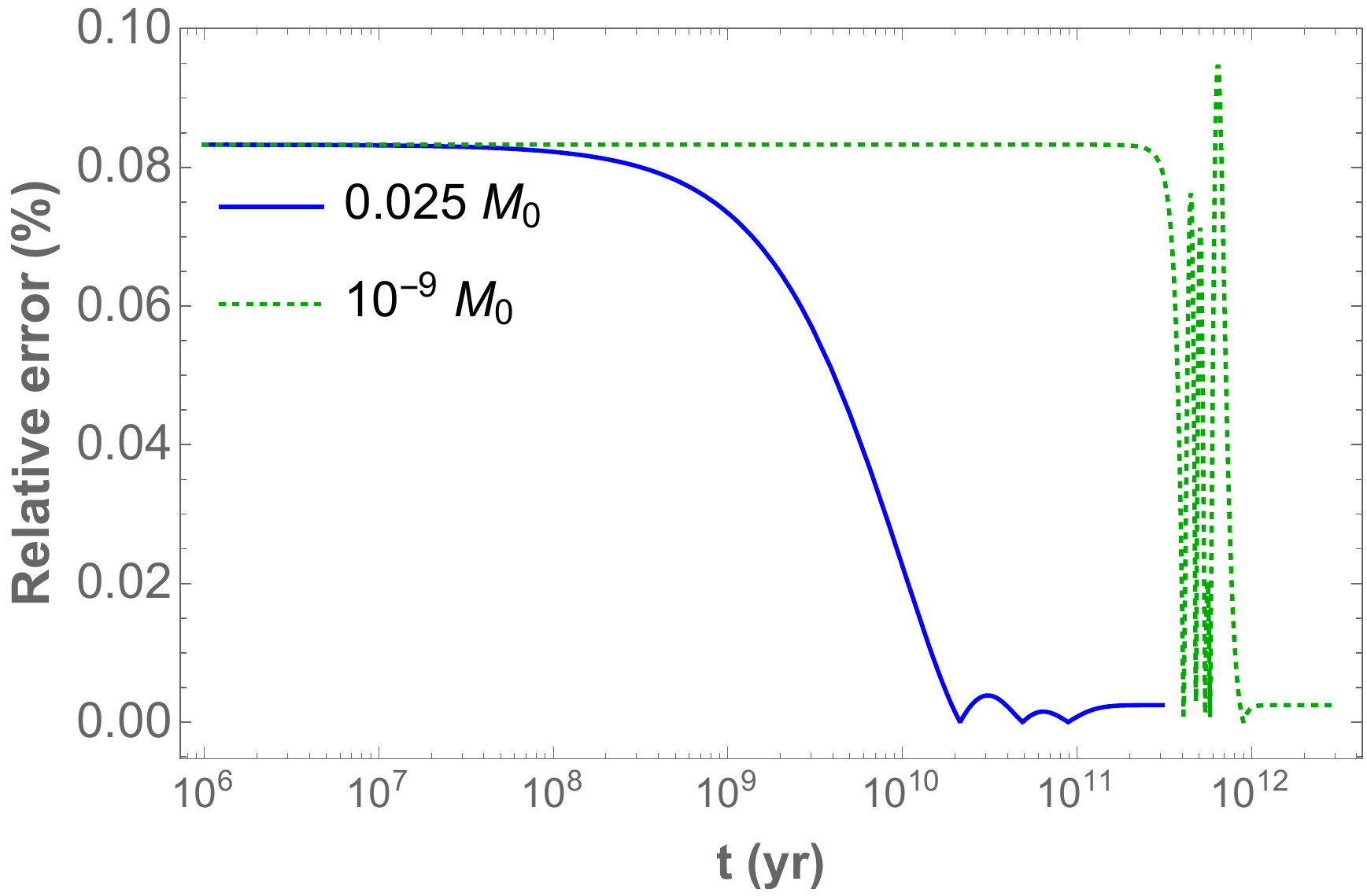}
		\caption{\label{fig:RelativeerrorshadowHeaviside}
			Relative error of the superradiant evolution of the shadow diameter $d_{\rm sh}$ of a black hole with initial mass $M_{0}=6.5\times10^{9}M_{\odot}$ and spin $\chi_{0}=0.8$ computed with the analytic approximation combined with the Heaviside tuning fit.
			We set the coupling $\alpha=0.05$ and consider scalar clouds with seed masses $M_{c,0}=0.025M_{0}$ (blue solid curve) and $M_{c,0}=10^{-9}M_{0}$ (green dotted curve). The end of the blue line corresponds to the end of the superradiant evolution.
		}
	\end{center}
\end{figure}

We can use this set of expressions to obtain a better agreement with the numerical data. For instance, Fig.~\ref{fig:RelativeerrorshadowHeaviside} shows the relative error of the shadow when using this fit.

\section{Shadow of the black hole}\label{app:shadowkerrbh}

Here, we derive approximate, analytic expressions to describe the dependence of the angular diameter $d_{\rm sh}$ on the 
black hole spin.

\subsection{Review}
\begin{figure}[htpb!]
\begin{center}
\includegraphics[width=0.5\textwidth,clip]{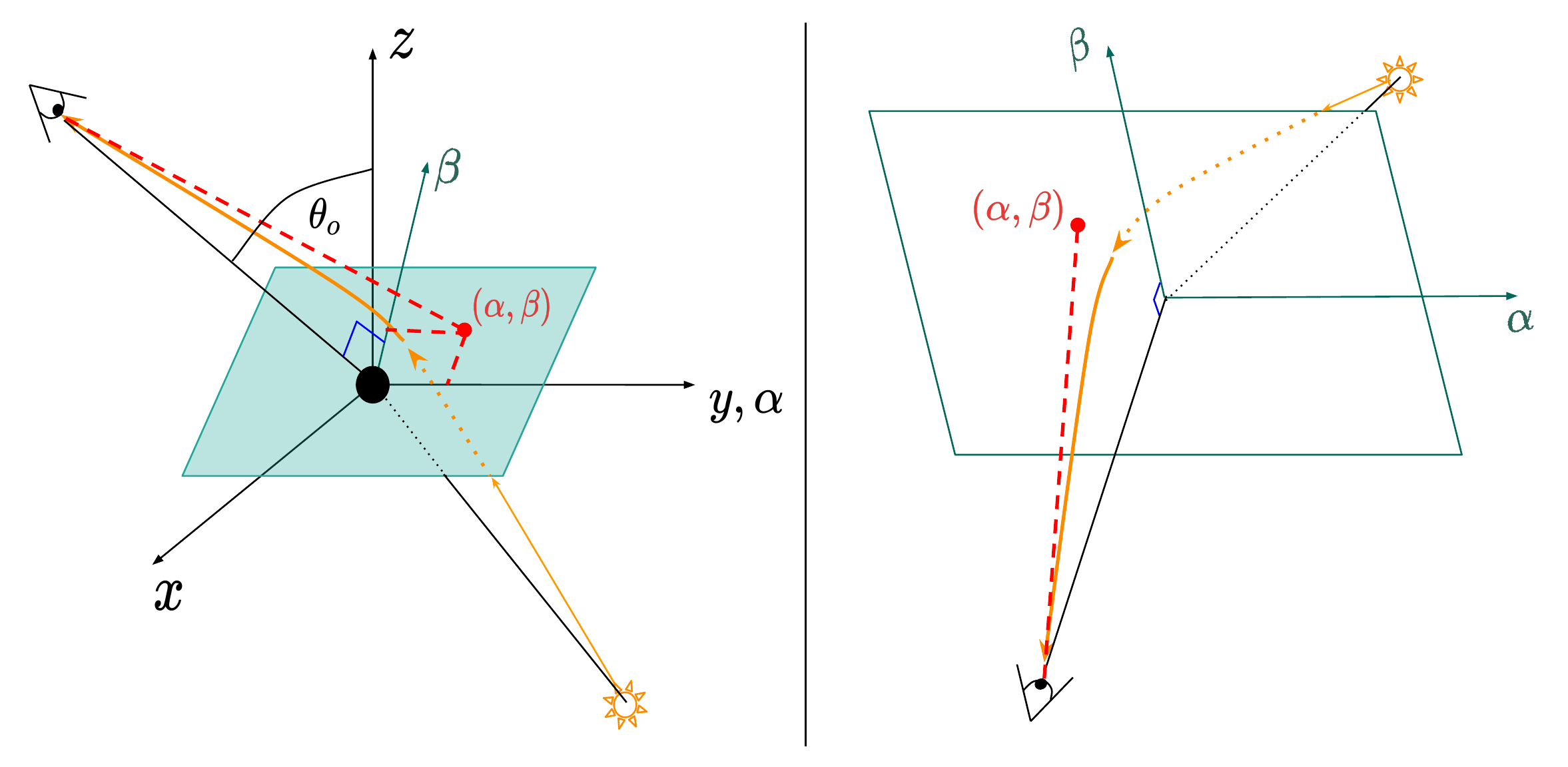}
\caption{\label{fig:ObsSky}
	Geometry of the system. The observer's sky plane is perpendicular to the line joining the observer and the black hole, situated at the origin and with the spin pointing in the z axis.}
\end{center}
\end{figure}

We consider a Kerr black hole of mass $M$ and angular momentum $J=a\,M = \chi\,M^{2}$~\footnote{Note, that here we derive the expressions for $a$, but
use the notation $\chi=a/M = J/M^2$ in the main text.}
given by the metric
\begin{align}
\label{appeq:kerrmetric}
\dif{s}^2 = &  -\left( 1 - \frac{2Mr}{\Sigma} \right) \dif t^2
        - \frac{4 a M r \sin^{2}\theta}{\Sigma} \dif t \dif \varphi
\nonumber \\ &
        + \frac{\Sigma}{\Delta} \dif r^{2} + \Sigma \dif\theta^{2}
        + \frac{\mathcal{F}}{\Sigma} \sin^{2}\theta \dif \varphi^{2}
\end{align}
in Boyer-Lindquist coordinates $(t,r,\theta,\varphi)$,
where the metric functions are
\begin{align*}
\Delta = & r^{2} + a^{2} - 2 M r
\,,\quad
\Sigma =r^{2} + a^{2}\cos^{2}\theta
\,,\\
\mathcal{F} = & \left(r^2 + a^2\right)^2 - \Delta a^2 \sin^2\theta
\,.
\end{align*}
In order to consider the case of a photon from infinity lensed by the Kerr black hole and reaching an observer at infinity, we need to study null geodesics. Null geodesics are described by the following set of differential equations from the integrals of motion of the Kerr black hole~\cite{PhysRev.174.1559}
\begin{align}
\Sigma{u^r}&=\pm\sqrt{R(r)}~,\label{appeq:uradial}\\
\Sigma{u^\theta}&=\pm\sqrt{\Theta(\theta)}~,\label{appeq:utheta}\\
\Sigma{u^\varphi}&=-\left(a-\frac{L}{\sin^2\theta}\right)+\frac{aP}{\Delta}~,\\
\Sigma{u^t}&=-a(a\sin^2\theta-L)+\frac{(r^2+a^2)P}{\Delta}\label{appeq:utime}~.
\end{align}
where 
$P = r^2 + a^2 - a L$, $u^{\mu}$ is the four-velocity of the photon, $L$ is the projection of the angular momentum of the photon onto the black hole's rotation axis,
$E$ is the energy of the photon and
\begin{align}
\label{appeq:PotentialR}
R(r) = & P^2-\Delta(Q+(L-aE)^2)
\,,\\
\label{appeq:PotentialTheta}
\Theta(\theta) = &Q-\cos^2\theta\left[-a^2E^2+L^2\sin^{-2}\theta\right]
\,.
\end{align}
$Q\equiv\kappa-(L-aE)^2$ is the Carter constant and $\kappa$ is a constant of separation used to solve the geodesics in the Hamilton-Jacobi framework \cite{PhysRev.174.1559}. 
In order to study what an observer would see, we define the observer's sky as the plane perpendicular to the line joining the observer and the black hole,
and determined by the coordinates $(\alpha,\beta)$,
as illustrated in Fig.~\ref{fig:ObsSky}.
Next, we derive a relation between the observer's sky coordinates $(\alpha,\beta)$ and the black hole's coordinates $(t,r,\theta,\varphi)$ which are defined
such that $z = r\cos\theta$ is aligned with the spin axis.
We furthermore define the observer's position angle $\theta_{o}$ between the observer's line of sight and the $z$ axis, where $\theta_{o}=0$ denotes an observer facing the
equatorial plane and $\theta_{o}=\pi/2$ corresponds to an observer lying in the equatorial plane.
We relate
\begin{align*}
\alpha = & -r_o^2\sin\theta_o\left.\frac{d\varphi}{dr}\right|_{r=r_o}
\,,\quad
\beta = r_o^2\left.\frac{d\theta}{dr}\right|_{r=r_o}&
\,.
\end{align*}
We note that $d\varphi/dr=u^\varphi/u^r$ and $d\theta/dr=u^\theta/u^r$,
employ Eqs.~\eqref{appeq:uradial}-\eqref{appeq:utime},
and take the limit $r_{o}\rightarrow\infty$ to find
\begin{align}
\alpha&=-\frac{L}{\sin\theta_o}~,\label{appeq:alphasky}\\
\beta&=\pm\sqrt{Q+a^2\cos^2\theta_o-L^2\cot^2\theta_o}~.\label{appeq:betasky}
\end{align}
Because the Kerr black hole is not spherically symmetric, 
the photons are not confined in a plane but acquire a precession movement. 
We define turning points by $u^\mu={dx^\mu}/{d\lambda}=0$, where $\lambda$ is the affine parameter. 
The radial turning point $r_{\text{min}}$ will be the largest positive root of $R(r)=0$, 
while the angular turning points $(\theta_{\text{min}}, \theta_{\text{max}})$ are the roots of $\Theta(\theta)=0$.

Since the geodesics are parametrized by $L$ and $Q$, it is useful to explore the region of the parameter space $(L,Q)$ at which photons that 
reach $r_{\text{min}}$ 
can escape to infinity \cite{Vazquez:2003zm}. 
We rewrite Eq.~\eqref{appeq:utheta} as
\begin{align*}
{u^{\theta}}^2 = & Q+a^2\cos^2\theta-L^2\cot^{2}\theta
\,.
\end{align*}
If we consider a photon crossing the equator $\theta=\pi/2$, we obtain $Q= {u^{\theta}}^2\geq0$. 
Taking into account photons that return to infinity means that $du^r/d\lambda>0$ at $r=r_{\text{min}}$, i.e., moving away from the black hole. Hence, in order to know the limiting case, we set $u^r|_{r=\bar{r}}=0$ and $du^r/d\lambda|_{r=\bar{r}}=0$, with $\bar{r}$ being the lower bound of $r_{\text{min}}$. With these conditions we are able to solve $Q$ and $L$ in terms of $\bar{r}$
\begin{align}
L(\bar{r})&=\frac{\bar{r}^2 (\bar{r}-3 M)+a^2(M+\bar{r})}{a (M-\bar{r})}\label{appeq:Lbar}~,\\
Q(\bar{r})&=\frac{\bar{r}^3 \left(4 a^2 M-\bar{r} (\bar{r}-3 M)^2\right)}{a^2 (M-\bar{r})^2}~.\label{appeq:Qbar}
\end{align}
The limiting case $Q=0$ yields two roots for $\bar{r}$ outside the event horizon namely $\bar{r}_+$ and $\bar{r}_-$. These roots are computed for a photon in the equator. In order to get some physical insight we can substitute these roots in $L(\bar{r})$, where we obtain $L(\bar{r}_+)>0$ and $L(\bar{r}_-)<0$, so we have two different turning points: $\bar{r}_+$ for photons moving with positive angular momentum, i.e., rotating in the same sense as the black hole; and $\bar{r}_-$ for photons moving with negative angular momentum, counterrotating with respect to the black hole \cite{Vazquez:2003zm}. From Eq.~\eqref{appeq:betasky} we can see that for photons to reach an observer, the argument inside the square root must be 
non-negative.
In the case of an observer in the equatorial plane the condition reduces to $Q\geq0$, so the values $r\in(\bar{r}_+,\bar{r}_-)$ correspond to photons reaching the equatorial plane from all different inclinations. In the case that the observer is not in the equatorial plane the condition for a photon to reach her/him is given by $Q+a^2\cos^2\theta-L^2\cot^2\theta\geq0$, 
with roots smaller than $[\bar{r}_+,\bar{r}_-]$.

\subsection{Black hole shadow formula derivation}\label{app:BHshadowapp}
Finding an exact formula of the angular diameter of the black hole shadow for all spins and inclinations is difficult due to 
the difficulty of finding the maximum approach distance $\bar{r}$, i.e. the inner edge of the shadow, for each inclination. 
We need to solve
\begin{align}\label{appeq:mastereqshadow}
Q+a^2\cos^2\theta-L^2\cot^2\theta = & 0
\,,
\end{align}
where $Q$ and $L$ are determined by Eqs.~\eqref{appeq:alphasky}, \eqref{appeq:Lbar} and \eqref{appeq:Qbar}.
In order to derive Eq. \eqref{eq:shadowdiameteranalytical} we have computed the angular diameter numerically for all inclinations. 
Because for inclinations different from 0 the shadow diameter is similar, we now fix $\theta_o=\pi/2$ for simplicity. 
For this orientation, Eq.~\eqref{appeq:mastereqshadow} implies $Q=0$.
Since our goal is to find an expression valid for all black hole mass and observer distance we propose the ansatz 
\begin{align}
\label{appeq:ansatzdshanalytic}
d_{sh} = & {A+B\left(1-\chi^2\right)^\delta}
\,,
\end{align}
where $\chi^{2}$ accounts for the symmetry $\chi\rightarrow-\chi$ and the exponent $\delta<1$ as indicated by our numerical computation.
To find the three unknown coefficients, we calculate the shadow diameter analytically
for a set of points in parameter space, namely $\chi=0$, $\chi=0.5$ and $\chi=1$.
For each of these points we determine $\bar{r}_{\pm}$ by determining the roots of Eq.~\eqref{appeq:Qbar} for $Q=0$,
and insert the result in~\eqref{appeq:Lbar} to calculate
\begin{align}
\label{appeq:dshadowcalc}
d_{\rm sh} = & \frac{1}{r_{o}} \left(|L_{-}| + |L_{+}| \right)
\,,
\end{align}
where $L_{\pm} = L (\bar{r}_{\pm})$.

{\noindent{\bf{Case $\chi=0$:}}
In this case we have to rederive Eqs.~\eqref{appeq:Lbar} and~\eqref{appeq:Qbar}, since their limit $\chi\equiv a/M=0$ is singular.
The radial potential~\eqref{appeq:PotentialR} becomes
\begin{align}
R(r)&=r^4-(r^2-2Mr)(Q+L^2)~.
\end{align}
Furthermore, $|L_{+}| = |L_{-}| = |L|$ due to symmetry, so the shadow diameter is given by
\begin{align}
d_{\rm sh}(\chi=0) = \frac{2 |L|}{r_{o}}
\,.
\end{align}
We use $Q=0$ and the condition that the photon arrives at the observer $du^r/d\lambda|_{r=\bar{r}}=0=u^r|_{r=\bar{r}}$, 
to find $\bar{r}=3M$ and $L_{\pm}=\pm3\sqrt{3}M$.
Inserting the result in the above expression gives
\begin{align}
\label{appeq:diametera=0}
d_{\rm sh}(\chi=0) = \frac{6\sqrt{3}M}{r_o}
\,,
\end{align}
which is the familiar expression for the shadow of a Schwarzschild black hole found in~\cite{Bozza:2009yw}.

{\noindent{\bf{Case $\chi=0.5$:}}
In this case we first solve Eq.~\eqref{appeq:Qbar}
to find the roots of $Q=0$. They are
$\bar{r}_{\pm} = M \left( 2+\cos\left(\pi/9\right)\pm\sqrt{3}\sin\left(\pi/9\right)\right)$.
Inserting this into Eq.~\eqref{appeq:Lbar} gives
\begin{align*}
L_{\pm} = & \frac{M}{2}\left[
        - 1 - 6 \cos\left(\frac{\pi}{9}\right) + 6 \cos\left(\frac{2\pi}{9}\right)
        \pm 6 \sqrt{3} S
\right]
\,,
\end{align*}
where we introduced $S = \sin\left(\pi/9\right) + \sin\left(2\pi / 9 \right)$.
Then, the shadow diameter~\eqref{appeq:dshadowcalc} is given by 
\begin{align}
\label{appeq:diametera=05}
d_{\rm sh} (\chi=0.5) = & \frac{6 \sqrt{3} M}{r_{o}} S
\,.
\end{align}

{\noindent{\bf{Case $\chi=1.0$:}}
As before, we solve for the roots of Eq.~\eqref{appeq:Qbar}, and find
$\bar{r}_{+}=4M$ and $\bar{r}_{-}=M$. Inserting the result into  Eqs.~\eqref{appeq:Qbar} and~\eqref{appeq:dshadowcalc}
gives
\begin{align}
\label{appeq:diametera=M}
d_{\rm sh} (\chi=1) = & \frac{9M}{r_{o}}
\,.
\end{align}

{\noindent{\bf{Determining the coefficients:}}}
We now insert our results~\eqref{appeq:diametera=0},~\eqref{appeq:diametera=05} and~\eqref{appeq:diametera=M}
into the ansatz~\eqref{appeq:ansatzdshanalytic} to identify the coefficients $(A,B,\delta)$.
We obtain
\begin{subequations}
\label{appeq:coefficients}
\begin{align}
A = & \frac{9M}{r_{o}}
\,,\\
B = & \frac{3 \left(2\sqrt{3}-3\right) M}{r_{o}}
\,,\\
\delta = & \frac{\ln(\frac{2\sqrt{3}-3}{2\sqrt{3}S-3})}{\ln(4/3)}
\,.
\end{align}
\end{subequations}

Finally, the dependence of the angular shadow diameter for an observer orientation
$\theta_o=\pi/2$
can be approximated by
\begin{align}
d_{\rm sh} = & \frac{3M}{r_o}\left[3+(2\sqrt{3}-3)\left(1-\chi^2\right)^\delta\right]
\,.
\end{align}

\bibliographystyle{apsrev4-1}
\bibliography{TFGbibliography.bib}

\end{document}